\documentclass[galaxies,article,referee,oneauthor,pdftex,12pt,a4paper]{mdpi}
\usepackage{graphicx}
\makeatletter
\renewcommand\@biblabel[1]{#1. }
\makeatother
\def\h0units{\mathrm{km\,s^{-1}\,Mpc^{-1}}}

\def\neave{<n_e>\;}

\newcommand{\om}{\Omega_{\rm M}}
\newcommand{\ok}{\Omega_K}
\newcommand{\ola}{\Omega_{\Lambda}}
\def\sun{\hbox{$\odot$}}
\newcommand{\mstar}{\ensuremath{m_{\text{B}}^\star}\,}
\def\aap{A\&A\,  }%% Astronomy and Astrophysics
\def\aj{AJ  }%% The Astronomical Journal
\def\apj{ApJ\,  }%% Astrophysical Journal
\def\apjs{ApJS  }%% Astrophysical Journal, Supplement
\def\jcap{Journal of Cosmology and Astroparticle Physic  } %Journal of Cosmology and Astroparticle Physics 
\def\jrasc{JRASC  } %journal Royal Astronomical Society Canada
\def\mnras{MNRAS\,  }%% Monthly Notices of the RAS

\def\pasa{PASA  }% % Publications of the Astronomical Society of Australia

\def\prd{Phys. Rev. D   }% % Physical Review D
\def\zap{Zeitschrift fur Astrophysik} %Zeitschrift fur Astrophysik
\lastpage{x}
\doinum{10.3390/------}
\pubvolume{xx}
\pubyear{2015}
\history{Received: xx / Accepted: xx / Published: xx}
\Title{
Pad\'e  Approximant and Minimax
Rational Approximation in Standard Cosmology
}

\Author{Lorenzo Zaninetti $^{1,}$}

\address
{
$^{1}$
Dipartimento  di Fisica, via P.Giuria 1,\\ I-10125 Turin,Italy
}

\corres
{
zaninetti@ph.unito.it
}

\abstract
{
The luminosity distance in the standard cosmology as given by
$\Lambda$CDM  and consequently the distance modulus for supernovae
can be defined  by
the Pad\'e  approximant.
A comparison with a known analytical
solution shows that
the Pad\'e  approximant  for the luminosity distance
has an error of $4\%$  at redshift $= 10$.
A similar procedure for the  Taylor expansion
of the luminosity distance
gives an error of $4\%$  at redshift $=0.7 $;
this means that for the luminosity distance,
the Pad\'e  approximation  is superior to the
Taylor series.
The availability of an analytical expression for the
distance modulus allows applying
the Levenberg--Marquardt  method to derive the fundamental
parameters from the available compilations for supernovae.
A new luminosity function for galaxies
derived from the truncated gamma
probability density function
models the observed luminosity
function for galaxies when the observed range
in absolute magnitude  is modeled  by the
Pad\'e  approximant.
A comparison of $\Lambda$CDM  with
other cosmologies is done
adopting a  statistical point of view.
}

\keyword
{
Cosmology;
Observational cosmology;
Distances, redshifts, radial velocities, spatial distribution of
galaxies;
Magnitudes and colors, luminosities
}

\PACS
{
98.80.-k  ;
98.80.Es
98.62.Py ;
98.62.Qz
}
\begin{document}

%%%%%%%%%%%%%%%%%%%%%%%%%%%%%%%%%%%%%%%%%%

\section{Introduction}

In order to obtain astronomical observables
such as  the distance modulus and  the absolute magnitude  for
supernovae (SN) of type Ia in the standard cosmological approach,
as given by the $\Lambda$CDM model,
we need  the evaluation
of the luminosity distance which is derived from  the comoving
distance.
At the moment of writing, there is no analytical expression
for the integral of the comoving distance in $\Lambda$CDM
and a numerical integration should be implemented.
An analytical expression for the integral of the
comoving distance in $\Lambda$CDM
can obtained by adopting the technique
of the Pad\'e  approximant, see \cite{2012Adachi,Aviles2014,Wei2014}.
Once  an approximate solution is obtained for the luminosity
distance  we can evaluate the
distance modulus and  the absolute magnitude for SNs.
Furthermore, the minimax
rational approximation  can provide a compact formula
for the two above astronomical observables as functions
of the redshift.
>From an observational point of view, the
progressive increase in the number of
supernova (SN) of type Ia
for which  the distance modulus is available,
34  SNe   in the sample which produced  evidence
for the   accelerating universe, see \cite{Riess1998},
580 SNe  in  the Union 2.1 compilation, see \cite{Suzuki2012}
and   740 SNe  in
the joint light-curve analysis (JLA), see \cite{Betoule2014},
allows analysing both the $\Lambda$CDM and other cosmologies from
a statistical point of view.
The statistical approach to cosmology is not new
and has been recently adopted by
\cite{Montiel2014} and \cite{Yahya2014}.
In order to cover the  previous  arguments,
Section \ref{secstandard} introduces
the Pad\'e  approximant and determines
the basic integral of the $\Lambda$CDM
which allows deriving the approximate luminosity distance.
The  approximate magnitude
here derived is applied  to parametrize
a new  luminosity function for galaxies
at high redshift, see Section \ref{sechigh}.
The distance modulus in different cosmologies
is reviewed and
the main  statistical parameters connected with the
distance modulus are derived,
see Section \ref{seccosmologies}.

\section{The standard cosmology}

This section introduces the Hubble distance, the dark energy
density, the curvature, the matter density, and  the comoving distance
(which is presented as the integral of the inverse of the
Hubble function).
In the absence of a general  analytical  formula for the comoving
distance, we introduce the Pad\'e approximation.
As a consequence, we deduce an approximate solution
for the transverse comoving distance, the luminosity distance,
and the distance modulus.
The shift that the Pad\'e approximation introduces
in the relationship for the poles is discussed.
The calibration of the Pad\'e approximation for the distance
modulus on two astronomical catalogs
allows deducing the minimax polynomial approximation
for the observed  distance modulus for SNs of type Ia.
\label{secstandard}

\subsection{The Pad\'e approximant}

We use the same symbols as in  \cite{Hogg1999},
where
the {\em Hubble
distance\/} $D_{\rm H}$
is defined as
\begin{equation}
\label{eq:dh}
D_{\rm H}\equiv\frac{c}{H_0}
\quad .
\end{equation}
We then introduce a first parameter
 $\om$
\begin{equation}
\om = \frac{8\pi\,G\,\rho_0}{3\,H_0^2}
\quad ,
\end{equation}
where $G$ is the Newtonian gravitational constant and
$\rho_0$ is the mass density at the present time.
A second parameter is $\ola$
\begin{equation}
\ola\equiv\frac{\Lambda\,c^2}{3\,H_0^2}
\quad ,
\end{equation}
where $\Lambda$ is the cosmological constant,
see \cite{Peebles1993}.
The two previous parameters are connected with the
curvature $\ok$ by
\begin{equation}
\om+\ola+\ok= 1
\quad .
\end{equation}
The  comoving distance, $D_{\rm C}$,  is
\begin{equation}
D_{\rm C} = D_{\rm H}\,\int_0^z\frac{dz'}{E(z')}
\label{integralez}
\end{equation}
where $E(z)$ is the `Hubble function'
\begin{equation}
\label{eq:ez}
E(z) = \sqrt{\om\,(1+z)^3+\ok\,(1+z)^2+\ola}
\quad .
\end{equation}
The above integral does not have an analytical
formula, except for the case of $\ola=0$, but the Pad\'e approximant, see
Appendix \ref{appendixb}, give an approximate evaluation
and the indefinite integral is (\ref{integral22})
where
the coefficients $a_j$ and $b_j$
can be found  in Appendix \ref{appendixa}.
The approximate definite integral for (\ref{integralez})
is therefore
\begin{equation}
D_{\rm C,2,2} = F_{2,2}(z;a_0,a_1,a_2,b_0,b_1,b_2)  -
F_{2,2}(0;a_0,a_1,a_2,b_0,b_1,b_2)
\label{integralez22}
\quad .
\end{equation}
The transverse comoving distance $D_{\rm M}$ is
\begin{equation}
D_{\rm M} = \left\{
\begin{array}{ll}
D_{\rm H}\,\frac{1}{\sqrt{\ok}}\,\sinh\left[\sqrt{\ok}\,D_{\rm C}/D_{\rm H}\right] & {\rm for}~\ok>0 \\
D_{\rm C} & {\rm for}~\ok=0 \\
D_{\rm H}\,\frac{1}{\sqrt{|\ok|}}\,\sin\left[\sqrt{|\ok|}\,D_{\rm C}/D_{\rm H}\right] & {\rm for}~\ok<0
\end{array}
\right.
\end{equation}
and the approximate transverse comoving distance $D_{\rm
M,2,2}$ computed with the Pad\'e approximant   is
\begin{equation}
D_{\rm M,2,2} = \left\{
\begin{array}{ll}
D_{\rm H}\,\frac{1}{\sqrt{\ok}}\,\sinh\left[\sqrt{\ok}\,
D_{\rm C,2,2}/D_{\rm H}\right] & {\rm for}~\ok>0 \\
D_{\rm C,2,2} & {\rm for}~\ok=0 \\
D_{\rm H}\,\frac{1}{\sqrt{|\ok|}}\,\sin\left[\sqrt{|\ok|}\,D_{\rm C,2,2}/D_{\rm H}\right] & {\rm for}~\ok<0
\end{array}
\right.
\end{equation}
An analytic expression for $D_{\rm M}$
can be obtained when
$\ola=0$:
\begin{equation}
D_{\rm M}=D_{\rm H}\,\frac{2\,[2-\om\,(1-z)-
(2-\om)\,\sqrt{1+\om\,z}]}{\om^2\,(1+z)}
~{\rm for}~\ola=0.
\end{equation}
This expression is useful for calibrating the numerical codes
which evaluate $D_{\rm M}$
when $\ola \neq 0$.

The luminosity distance is
\begin{equation}
D_{\rm L} = (1+z)\,D_{\rm M}
\label{luminositydistance}
\end{equation}
which in the case
of
$\ola=0$
becomes
\begin{equation}
D_{\rm L} =
2\,{\frac {c \left( 2-{\it \om}\, \left( 1-z \right) - \left( 2-{
\it \om} \right) \sqrt {z{\it \om}+1} \right) }{H_{{0}}{{\it
\om}}^{2}}}
\quad ,
\label{relativisticlumdist}
\end{equation}
and  the distance modulus when $\ola=0$ is
\begin{equation}
m-M =
25+5\,{\frac {1}{\ln  \bigl( 10 \bigr) }\ln  \bigl( 2\,{\frac {c
 \bigl( 2-{\it \om}\, \bigl( 1-z \bigr) - \bigl( 2-{\it \om}
 \bigr) \sqrt {z{\it \om}+1} \bigr) }{H_{{0}}{{\it \om}}^{2}}}
 \bigr) }
\quad  .
\label{modulusrelativistic}
\end{equation}
The Pad\'e approximant  luminosity distance
when  $\ola \neq 0$
 is
\begin{equation}
D_{\rm L,2,2} = (1+z)\,D_{\rm M,2,2}
\label{luminositydistancepade}
\quad ,
\end{equation}
and the Pad\'e approximant distance modulus, $(m-M)_{2,2}$,
in its compact version,
is
\begin{equation}
(m-M)_{2,2} =25 +5 \log_{10}(D_{\rm L,2,2})
\quad ,
\end{equation}
and, as a consequence, the  Pad\'e approximant  absolute
magnitude, $M_{2,2}$, is
\begin{equation}
M_{2,2} = m -25 -5 \log_{10}(D_{\rm L,2,2})
\quad .
\label{absmagz}
\end{equation}

The expanded  version of the Pad\'e
approximant distance modulus
is
\begin{equation}
(m-M)_{2,2} =
25+5\,{\frac {1}{\ln  \left( 10 \right) }\ln  \left( {\frac {c \left(
1+z \right) }{H_{{0}}\sqrt {{\it \ok}}}\sinh \left( 1/2\,{\frac {
\sqrt {{\it \ok}}A}{{b_{{2}}}^{2}\sqrt {4\,b_{{0}}b_{{2}}-{b_{{1}}}
^{2}}}} \right) } \right) }
\quad ,
\label{distancemodulusexplicit}
\end{equation}
with
\begin{eqnarray}
A=
\ln  \left( {z}^{2}b_{{2}}+zb_{{1}}+b_{{0}} \right) a_{{1}}b_{{2}}
\sqrt {4\,b_{{0}}b_{{2}}-{b_{{1}}}^{2}}-\ln  \left( {z}^{2}b_{{2}}+zb_
{{1}}+b_{{0}} \right) a_{{2}}b_{{1}}\sqrt {4\,b_{{0}}b_{{2}}-{b_{{1}}}
^{2}}
\nonumber \\
-\ln  \left( b_{{0}} \right) a_{{1}}b_{{2}}\sqrt {4\,b_{{0}}b_{{2
}}-{b_{{1}}}^{2}}+\ln  \left( b_{{0}} \right) a_{{2}}b_{{1}}\sqrt {4\,
b_{{0}}b_{{2}}-{b_{{1}}}^{2}}+2\,a_{{2}}zb_{{2}}\sqrt {4\,b_{{0}}b_{{2
}}-{b_{{1}}}^{2}}
\nonumber \\
+4\,\arctan \left( {\frac {2\,zb_{{2}}+b_{{1}}}{
\sqrt {4\,b_{{0}}b_{{2}}-{b_{{1}}}^{2}}}} \right) a_{{0}}{b_{{2}}}^{2}
-2\,\arctan \left( {\frac {2\,zb_{{2}}+b_{{1}}}{\sqrt {4\,b_{{0}}b_{{2
}}-{b_{{1}}}^{2}}}} \right) b_{{1}}a_{{1}}b_{{2}}
\nonumber \\
-4\,\arctan \left( {
\frac {2\,zb_{{2}}+b_{{1}}}{\sqrt {4\,b_{{0}}b_{{2}}-{b_{{1}}}^{2}}}}
 \right) a_{{2}}b_{{0}}b_{{2}}+2\,\arctan \left( {\frac {2\,zb_{{2}}+b
_{{1}}}{\sqrt {4\,b_{{0}}b_{{2}}-{b_{{1}}}^{2}}}} \right) {b_{{1}}}^{2
}a_{{2}}
\nonumber \\
-4\,\arctan \left( {\frac {b_{{1}}}{\sqrt {4\,b_{{0}}b_{{2}}-{
b_{{1}}}^{2}}}} \right) a_{{0}}{b_{{2}}}^{2}+2\,\arctan \left( {\frac
{b_{{1}}}{\sqrt {4\,b_{{0}}b_{{2}}-{b_{{1}}}^{2}}}} \right) b_{{1}}a_{
{1}}b_{{2}}
\nonumber \\
+4\,\arctan \left( {\frac {b_{{1}}}{\sqrt {4\,b_{{0}}b_{{2}
}-{b_{{1}}}^{2}}}} \right) a_{{2}}b_{{0}}b_{{2}}-2\,\arctan \left( {
\frac {b_{{1}}}{\sqrt {4\,b_{{0}}b_{{2}}-{b_{{1}}}^{2}}}} \right) {b_{
{1}}}^{2}a_{{2}}
\nonumber
\end{eqnarray}
The above procedure can also be applied when
the argument of the integral (\ref{integralez})
is expanded about z=0 in a Taylor series of order 6.
The resulting luminosity distance, $D_{\rm L,6}$,
is
\begin{equation}
D_{\rm L,6}= -{\frac {c \left( 1+z \right) }{\sqrt {{\it \ok}}H_{{0}}}}
\sinh \left( {\frac {\sqrt {{\it \ok}}z{\it C_T}}{7680}} \right)
\label{luminositydistancetaylor}
\end{equation}
where
\begin{eqnarray}
C_T=315\,{{\it \om}}^{5}{z}^{5}+350\,{{\it \om}}^{4}{z}^{5}-420\,{{
\it \om}}^{4}{z}^{4}+400\,{{\it \om}}^{3}{z}^{5}-480\,{{\it
\om}}^{3}{z}^{4}+480\,{{\it \om}}^{2}{z}^{5}
\nonumber \\
+600\,{{\it \om}}
^{3}{z}^{3}-576\,{{\it \om}}^{2}{z}^{4}+640\,{z}^{5}{\it \om}+
720\,{{\it \om}}^{2}{z}^{3}-768\,{z}^{4}{\it \om}+1280\,{z}^{5}-
960\,{{\it \om}}^{2}{z}^{2}
\nonumber \\
+960\,{z}^{3}{\it \om}-1536\,{z}^{4}-
1280\,{z}^{2}{\it \om}+1920\,{z}^{3}+1920\,z{\it \om}-2560\,{z}^
{2}+3840\,z-7680
\end{eqnarray}
The    goodness of the approximation is evaluated
through the percentage error, $\delta$, which is
\begin{equation}
\delta = \frac{\big | D_{\rm L}(z) - D_{\rm L,app}(z) \big |}
{D_{\rm L}(z)} \times 100
\quad ,
\end{equation}
where $D_{\rm L}(z)$ is the exact luminosity distance when
$\ola=0$,  see Eqn.~(\ref{luminositydistance})
and  $D_{\rm L,app}(z)$ is the Taylor or Pad\'e  approximate
luminosity distance, see also formula (2.12) in \cite{2012Adachi}.

% figure   deltataylor
\begin{figure}
\begin{center}
\includegraphics[width=10cm]{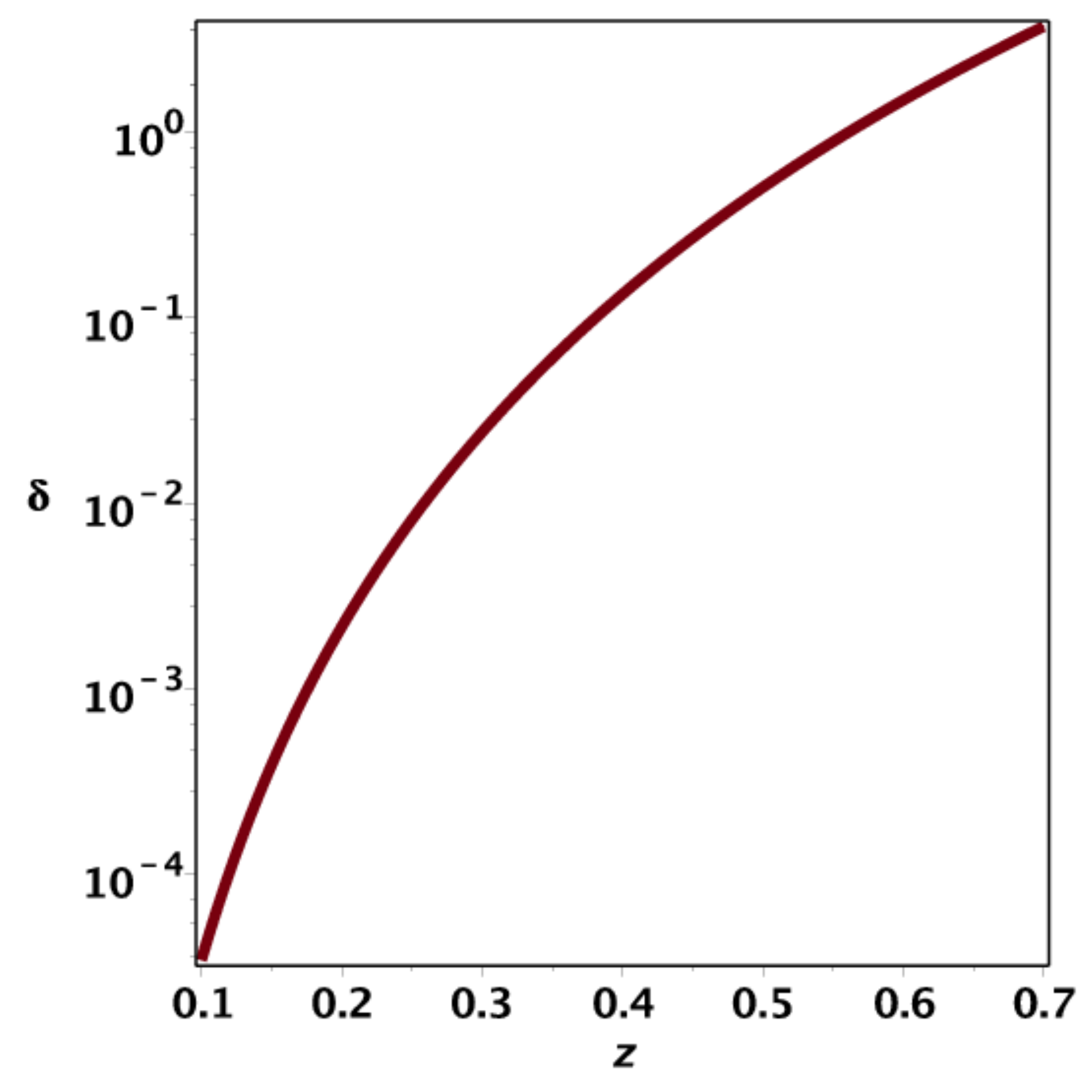}
\end{center}
\caption
{
Percentage error, $\delta$,
relative to the Taylor approximated luminosity distance,
 see Eq.~(\ref{luminositydistancetaylor}),
when $H_0 = 69.6 \h0units$  and $\om=0.9$.
}
\label{deltataylor}
\end{figure}
% end deltataylor

% figure   deltapade
\begin{figure}
\begin{center}
\includegraphics[width=10cm]{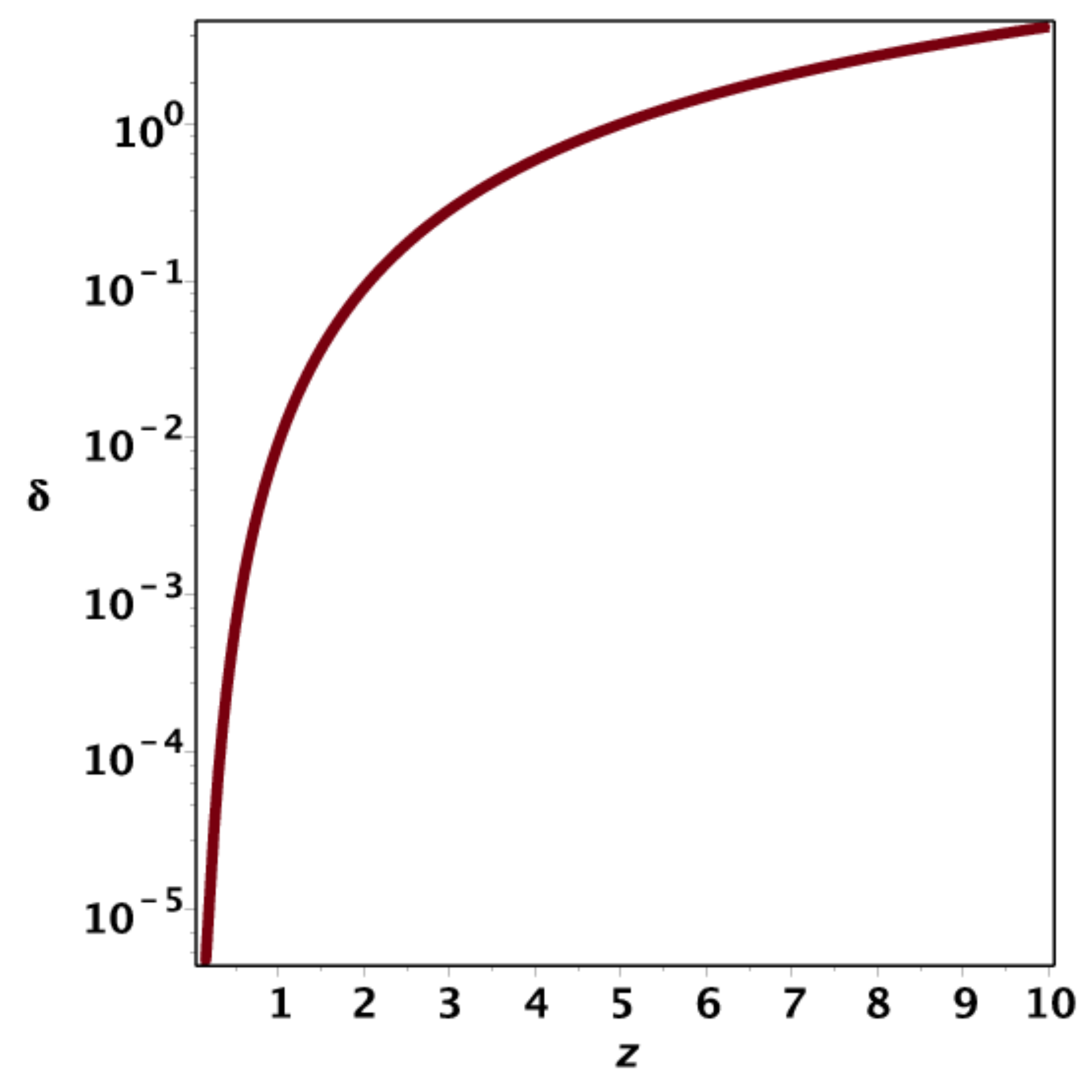}
\end{center}
\caption
{
Percentage error, $\delta$,
relative to the Pad{\`e}  approximated luminosity distance,
see Eq.~(\ref{luminositydistancepade}),
when $H_0 = 69.6 \h0units$  and $\om=0.9$.
}
\label{deltapade}
\end{figure}
% end deltapade

Figures \ref{deltataylor} and \ref{deltapade} report
%citiamofigura_deltataylor
%citiamofigura_deltapade
the percentage error as a function of the redshift $z$
for the Taylor and Pad\'e approximations, respectively.
The Pad\'e approximation is superior
to the truncated Taylor expansion because $\delta \approx 4$
is reached at $z=10$
for the Pad\'e approximant   and at $z=0.7$ for the Taylor expansion.

\subsection{The presence of poles}

\begin{figure}
\begin{center}
\includegraphics[width=10cm]{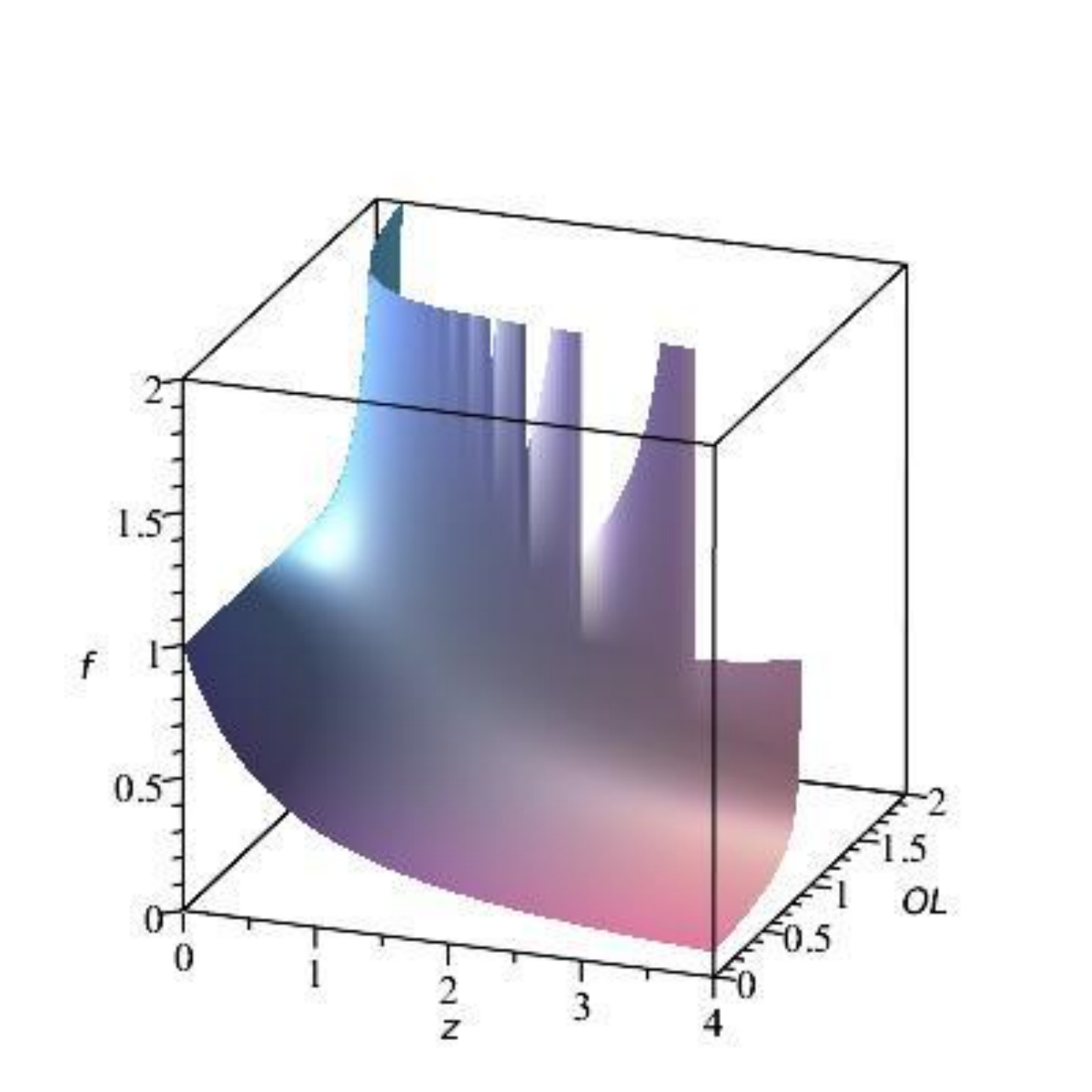}
\end{center}
\caption
{
Behavior of $\frac{1}{E(z)}$
as a function of $z$ and $\ola$ in the neighbourhoods  of the poles
when $\ok=0.11$.
}
\label{polessurface}
\end{figure}

The integrand of (\ref{integralez}) contains poles or singularities for a given set of parameters, see Figure \ref{polessurface}.
% figure   polessurface
% end polessurface

\begin{figure}
\begin{center}
\includegraphics[width=10cm]{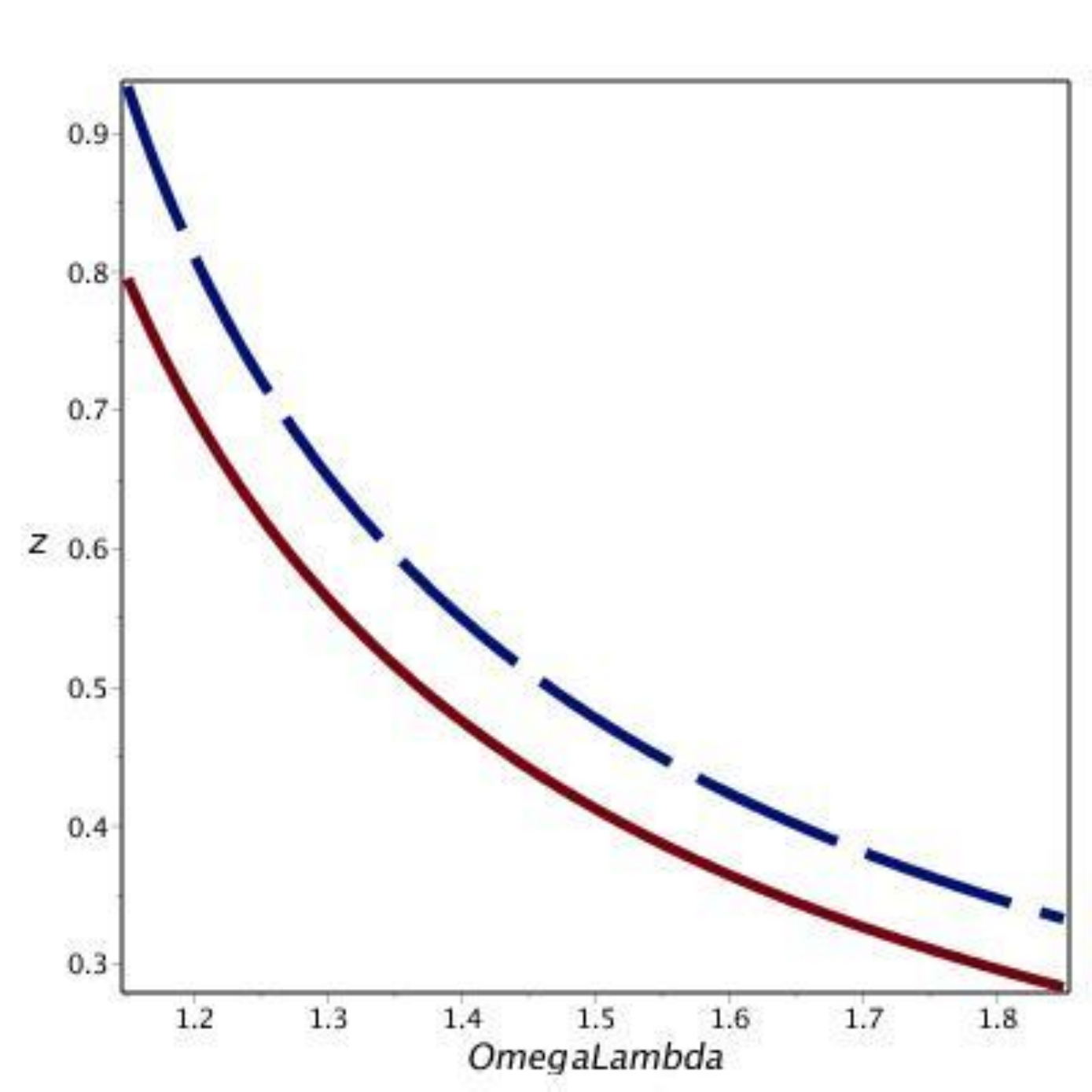}
\end{center}
\caption
{
The exact solution for the zero in E(z), full red line,
and Pad\'e approximated solution, dashed blue line,
when $\ok=0.11$.
}
\label{polespade}
\end{figure}

The equation which models the poles is
\begin{equation}
E(z) = 0.
\end{equation}
The exact solution of the above equation $z(\ola;\ok=0.11)$
is shown in Figure \ref{polespade} together
with the Pad\'e approximated solution $z_{2,2}(\ola;\ok=0.11)$.
% figure   polespade
% end polespade
Is therefore possible to conclude
that the Pad\'e approximation
shifts
the locations of the poles by
$\Delta z$; this shift  expressed as
a percentage error is $\delta \approx 17\%$ in the considered interval
$\ola=[1.15,1.85]$.

\subsection{An astrophysical application}

We now have  a Pad\'e   approximant  expression
for the distance modulus as a function of
of $H_0$, $\om$ and $\ola$.
We now perform an astronomical test
on the 580 SNe  in  the Union 2.1 compilation, see \cite{Suzuki2012}
and  on the  740 SNe  in
the joint light-curve analysis (JLA).
The JLA  compilation is available at
the Strasbourg Astronomical Data Center (CDS)
and consists of  SNe (type I-a)  for which
we have  a heliocentric redshift, $z$, apparent
magnitude $\mstar$ in the B band, error in $\mstar$, $\sigma_{\mstar}$,
parameter $X1$,   error in $X1$,
$\sigma_{X1}$,
parameter $C$, error in the parameter $C$, $\sigma_C$  and
$\log_{10} (M_{stellar})$.
The observed distance modulus  is defined by
Eq.~(4) in \cite{Betoule2014}
\begin{equation}
m-M =
-C\beta+{\it X1}\,\alpha-M_{{b}}+ \mstar
\quad.
\end{equation}
The  adopted parameters are
$\alpha=0.141$, $\beta=3.101$   and
\begin{equation}
M_{{b}} = \begin{cases}
-19.05  & \text{if } M_{stellar} <    10^{10} M_{\sun} \\
-19.12  & \text{if } M_{stellar} \geq 10^{10} M_{\sun}
 \end{cases}
\quad ,
\end{equation}
where $M_{\sun}$ is the mass of the sun,
see line 1  in Table 10 of
\cite{Betoule2014}.
The uncertainty  in the observed distance modulus,
$\sigma_{m-M}$,
is  found by implementing the error
propagation equation (often called the law of errors of Gauss) when
the covariant terms are neglected, see  equation (3.14)
in \cite{bevington2003},
\begin{equation}
\sigma_{m-M} =
\sqrt {{\alpha}^{2}{\sigma_{{{\it X1}}}}^{2}+{\beta}^{2}{\sigma_{{C}}}
^{2}+{\sigma_{{{\it \mstar}}}}^{2}}
\quad .
\end{equation}
The three astronomical parameters in question,
$H_0$, $\om$ and $\ola$,
can be derived trough the
Levenberg--Marquardt  method
(subroutine MRQMIN in \cite{press})
once  an analytical expression
for the  derivatives  of the distance modulus
with respect to the unknown  parameters is  provided.
As a practical example, the
derivative  of the distance modulus, $(m-M)_{2,2}$,
with respect to $H_0$ is
\begin{equation}
\frac{d(m-M)_{2,2}}{dH_0}=-5\,{\frac {1}{H_{{0}}\ln  \left( 10 \right) }}
\quad .
\end{equation}

\begin{figure}
\begin{center}
\includegraphics[width=10cm]{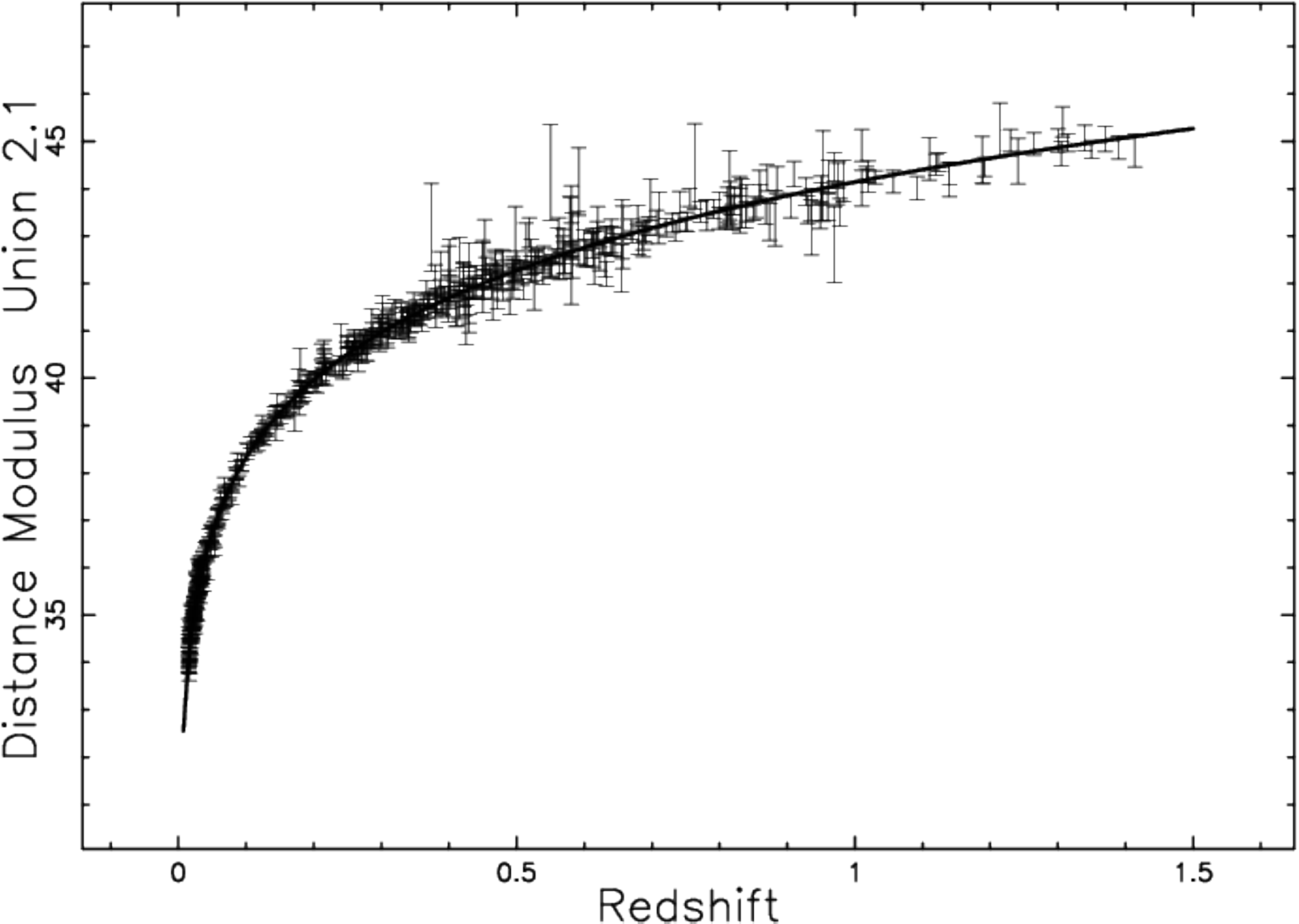}
\end{center}
\caption{
Hubble diagram for the  Union 2.1 compilation.
The solid line represents the best fit
for the approximate distance modulus  as represented by
Eq. (\ref{distancemodulusexplicit}),
parameters as in Table \ref{chi2value}.
}
\label{distmodpade}
\end{figure}

\begin{figure}
\begin{center}
\includegraphics[width=10cm]{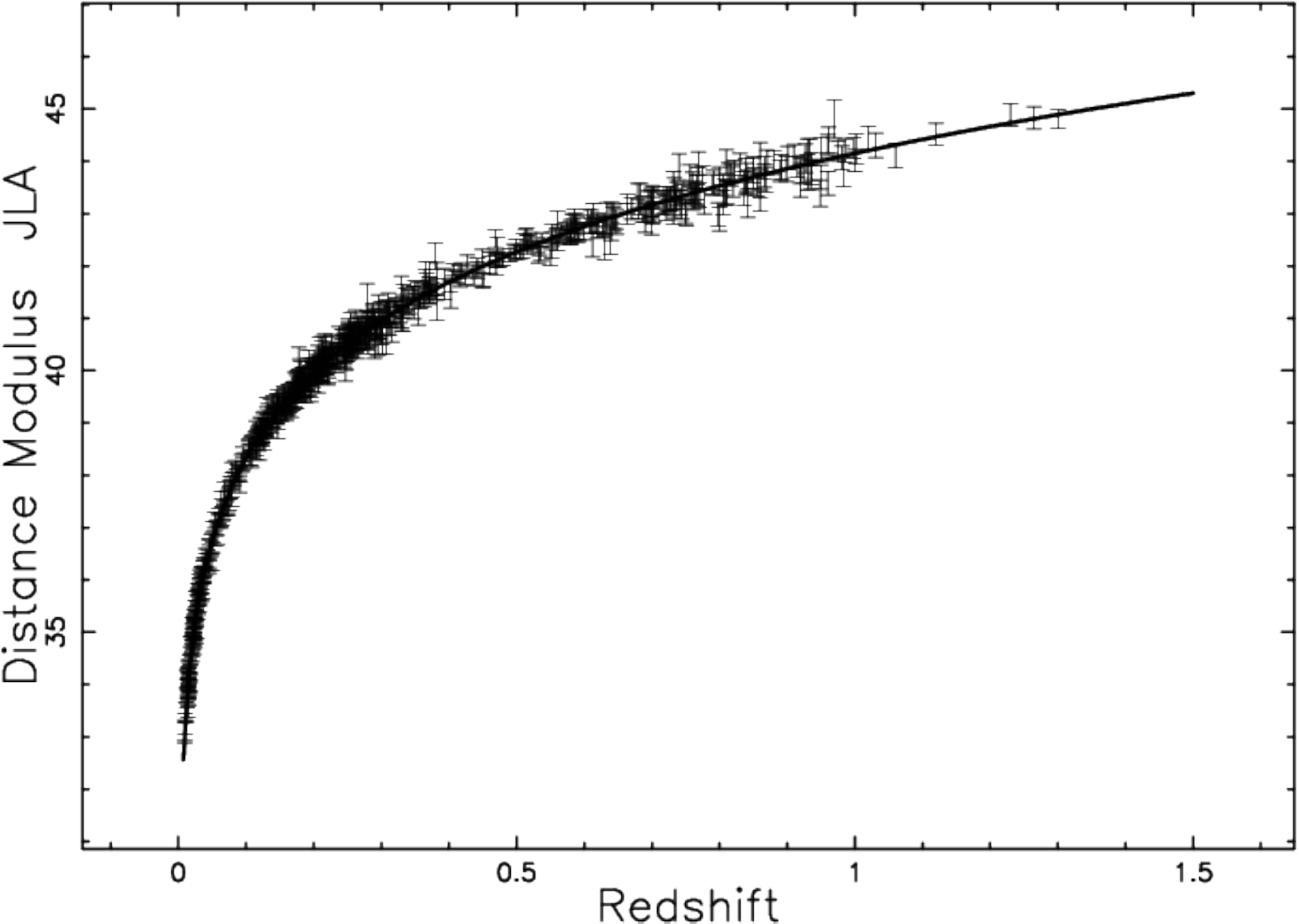}
\end{center}
\caption{
Hubble diagram for the  JLA compilation.
The solid line represents the best fit
for the approximate distance modulus  as given by
Eq. (\ref{distancemodulusexplicit}),
parameters as in Table \ref{chi2value}.
}
\label{distmodpade_jla}
\end{figure}

This numerical procedure minimizes the merit function $\chi^2$
evaluated as
\begin{equation}
\chi^2  = \sum_{i=1}^N \biggr [\frac{(m-M)_i - (m-M)(z_i)_{th}}{\sigma_i}\biggl] ^2
\quad ,
\label{chisquare}
\end{equation}
where $N=480$, $(m-M)_i$ is the observed distance modulus
evaluated at $z_i$,
$\sigma_i$ is the error in the observed distance
modulus evaluated at $z_i$,
and
$(m-M)(z_i)_{th}$ is the theoretical distance modulus evaluated at
$z_i$, see formula (15.5.5) in \cite{press}.
A reduced  merit function $\chi_{red}^2$
is  evaluated  by
\begin{equation}
\chi_{red}^2 = \chi^2/NF
\quad,
\label{chisquarereduced}
\end{equation}
where $NF=n-k$ is the number of degrees  of freedom,
       $n$     is the number of SNe,
and    $k$     is the number of parameters.
Another useful statistical parameter is  the associated $Q$-value,
which has   to be understood as the
 maximum probability of obtaining a better fitting,
 see formula (15.2.12) in \cite {press}:
\begin{equation}
Q=1- GAMMQ (\frac{N-k}{2},\frac{\chi^2}{2} )
\quad ,
\end{equation}
where GAMMQ is a subroutine  for the incomplete gamma function.
The Akaike information criterion
(AIC), see \cite{Akaike1974},
is defined by
\begin{equation}
AIC  = 2k - 2  ln(L)
\quad,
\end{equation}
where $L$ is
the likelihood  function.
We assume  a Gaussian distribution for  the errors
and  the likelihood  function
can be derived  from the $\chi^2$ statistic
$L \propto \exp (- \frac{\chi^2}{2} ) $
where  $\chi^2$ has been computed by
Eq.~(\ref{chisquare}),
see~\cite{Liddle2004}, \cite{Godlowski2005}.
Now the AIC becomes
\begin{equation}
AIC  = 2k + \chi^2
\quad .
\label{AIC}
\end{equation}
Table \ref{chi2value} reports the three astronomical parameters
for the two catalogs of SNs and
Figures \ref{distmodpade} and \ref{distmodpade_jla}
%citiamofigura_distmodpade
%citiamofigura_distmodpade_jla
display the best  fits.

\begin{table}[ht!]
\caption
{
Numerical values of  $\chi^2$, $\chi_{red}^2$, $Q$,
and the AIC of the Hubble diagram for two compilations,
$k$ stands for the number of parameters.
}
\label{chi2value}
\begin{center}
\begin{tabular}{|c|c|c|c|c|c|c|c|}
\hline
compilation   &  SNs& k    &   parameters    & $\chi2$& $\chi_{red}^2$
&
Q  & AIC    \\
\hline
Union~2.1 & 577 &  3
& $H_0$ = 69.81; $\om=0.239$; $\ola=0.651$
& 562.699 &  0.975  & 0.657 &   568.699  \\
\hline
JLA  & 740 &  3
& $H_0$ = 69.398; $\om=0.181$; $\ola=0.538$
& 625.733 &  0.849  & 0.998  &   631.733  \\
\hline
\end{tabular}
\end{center}
\end{table}

% figure   distmodpade
% end distmodpade

% figure   distmodpade_jla
\begin{figure}
\begin{center}
\includegraphics[width=10cm]{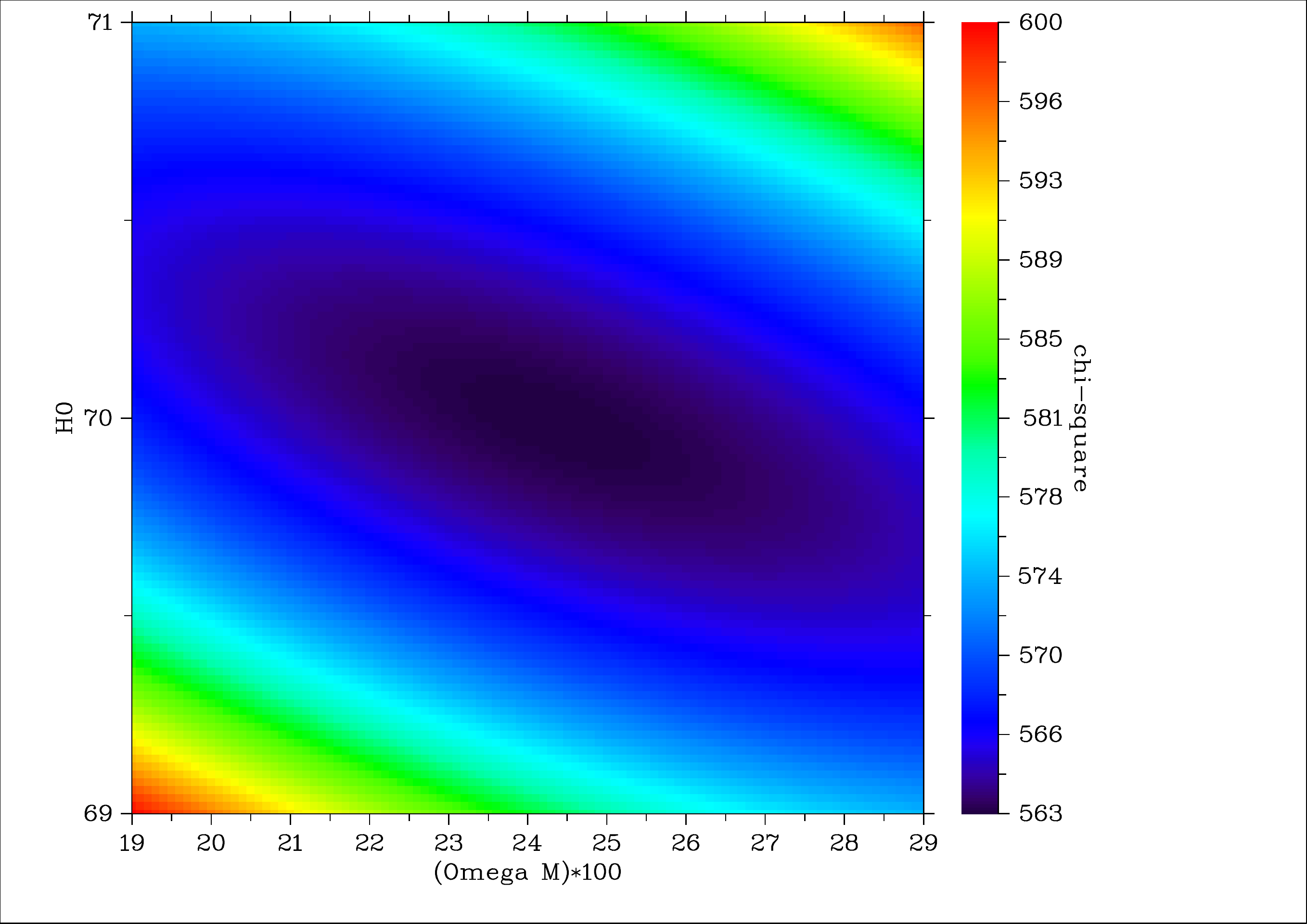}
\end{center}
\caption
{
Color contour plot for $\chi2$  of  the Hubble diagram for the Union 2.1 compilation when $H_0$ and $\om$ are variables and
$\ola=0.651$.
}
\label{chi2_gnuplot}
\end{figure}

% end distmodpade_jla
In order to see how $\chi2$ varies around the minimum
found by the Levenberg--Marquardt method, Figure
\ref{chi2_gnuplot}
%citiamofigura_chi2_gnuplot
presents a 2D color map for the
values of $\chi2$  when $H_0$ and $\om$ are allowed
to vary around the numerical values which  fix the minimum.
% figure   chi2_gnuplot
% end chi2_gnuplot

The Pad\'e approximant  distance modulus
has a simple expression when the
minimax rational approximation is used,
as an example $p=3,q=2$,
see Appendix  \ref{appendixc} for the meaning of $p$ and $q$.
In the case of the Union 2.1 compilation,
the approximation of formula (\ref{distancemodulusexplicit})
with the parameters of Table  \ref{chi2value}
over the range in $z\in [0,4]$ gives
the following minimax equation
\begin{equation}
(m-M)_{3,2} =
\frac
{
 0.359725+ 5.612031\,z+ 5.627811\,{z}^{2}+ 0.054794\,{z
}^{3}
}
{
0.010587+ 0.137541\,z+ 0.115904\,{z}^{2}
}
 \quad Union~2.1~compilation
\quad ,
\label{distmodminimaxunion21}
\end{equation}
the maximum error being 0.0024.
The maximum error of the polynomial approximation
as a function of $p$ and $q$ is shown in Table \ref{tableminimax}.

\begin{table}
\caption { The maximum error in the minimax rational approximation
 for the distance modulus in the case of the Union 2.1 compilation.  }
 \label{tableminimax}
 \[
 \begin{array}{ccc}
 \hline
 \hline
 \noalign{\smallskip}
$p$  &   $q$   & maximum~ error       \\
 \noalign{\smallskip}
 \hline
\noalign{\smallskip}
1 & 1 & 0.2872 \\
2 & 2 & 0.0197 \\
3 & 2 & 0.0024 \\
3 & 3 & 0.0006 \\
\noalign{\smallskip}
\noalign{\smallskip}
 \hline
 \hline
 \end{array}
 \]
 \end {table}

In the case of the JLA compilation, the minimax equation is
\begin{equation}
(m-M)_{3,2} =
\frac
{
 0.442988+ 6.355991\,z+ 5.40531\,{z}^{2}+ 0.044133\,{z
}^{3}
}
{
 0.012985+ 0.154698\,z+ 0.109749\,{z}^{2}
}
\quad JLA~compilation
\quad  ,
\label{distmodminimaxjla}
\end{equation}
the maximum error being 0.003.

The maximum difference between  the two minimax formulas
which approximate the distance modulus,
Eqs.~(\ref{distmodminimaxunion21}) and (\ref{distmodminimaxjla}),
is at $z=4$, and is 0.0584  mag.
In the case of the luminosity distance
as given by the Pad\'e approximation, see Eq.~(\ref{luminositydistancepade}),
 the minimax approximation
gives
\begin{subequations}
\begin{align}
D_{\rm L,3,2}=&
\frac
{
- 7.7618- 1788.535\,z- 3203.0635\,{z}^{2}- 65.8463\,{z}^{3}
}
{
- 0.438- 0.3348\,z+ 0.02039\,{z}^{2}
}\,Mpc
&Union~2.1\\
\label{dlzminimaxjla}
D_{\rm L,3,2}=&
\frac
{
- 1.1674- 2413.8956\,z- 2831.4248\,{z}^{2}- 100.2959\,{z}^{3}
}
{
- 0.562- 0.2367\,z+ 0.007746\,{z}^{2}
}\,Mpc
&JLA
\end{align}
\end{subequations}

\section{Application at high redshift}

\label{sechigh}
This section introduces a new luminosity function (LF)
for galaxies, which has a lower and an upper bound.
The presence of a lower bound for the
luminosity of galaxies  allows to model the evolution
of the LF as a function of the redshift.

\subsection{The Schechter luminosity function}

The  Schechter LF,
after \cite{schechter},
is the standard LF for galaxies:
\begin{equation}
\Phi (\frac{L}{L^*}) dL  = (\frac {\Phi^*}{L^*}) (\frac {L}{L^*})^{\alpha}
\exp \bigl ( {-  \frac {L}{L^*}} \bigr ) dL.
\label{equation_schechter}
\end{equation}
Here, $\alpha$ sets the shape,
$L^*$ is the
characteristic luminosity,  and $\Phi^*$ is the normalization.
The  distribution in absolute magnitude is
\begin{equation}
\Phi (M)dM=0.921 \Phi^* 10^{0.4(\alpha +1 ) (M^*-M)}
\exp \bigl ({- 10^{0.4(M^*-M)}} \bigr)  dM \, ,
\label{lfstandard}
\end{equation}
where $M^*$ is the characteristic magnitude.

 \subsection{The gamma luminosity function}

 The {\em gamma } LF
  is
\begin {equation}
f(L;\Psi^*,L^*,c) = \Psi^*
\frac {
 \left( {\frac {L}{L^*}} \right) ^{c-1}{{\rm e}^{-{\frac {L}{L^*}}}}
}
{
L^*\Gamma  \left( c \right)
}
\label{gammastandard}
\end {equation}
where
$\Psi^*$ is the total number of galaxies per unit Mpc$^3$,
\begin{equation}
\mathop{\Gamma\/}\nolimits\!\left(z\right)
=\int_{0}^{\infty}e^{{-t}}t^{{z-1}}dt
\quad ,
\end{equation}
is the gamma function,
    $L^*>0$   is the scale
and $c>0$    is the shape,
see  formula (17.23) in \cite{univariate1}.
Its expected value   is
\begin{equation}
E(\Psi^*,L^*,c)= \Psi^* L^*c  \quad.
\end{equation}
The change of parameter $(c-1)=\alpha$ allows obtaining
the same scaling as for the
Schechter LF (\ref{equation_schechter}).

 \subsection{The truncated gamma luminosity function}

\begin{figure}
 \centering
\includegraphics[width=10cm]{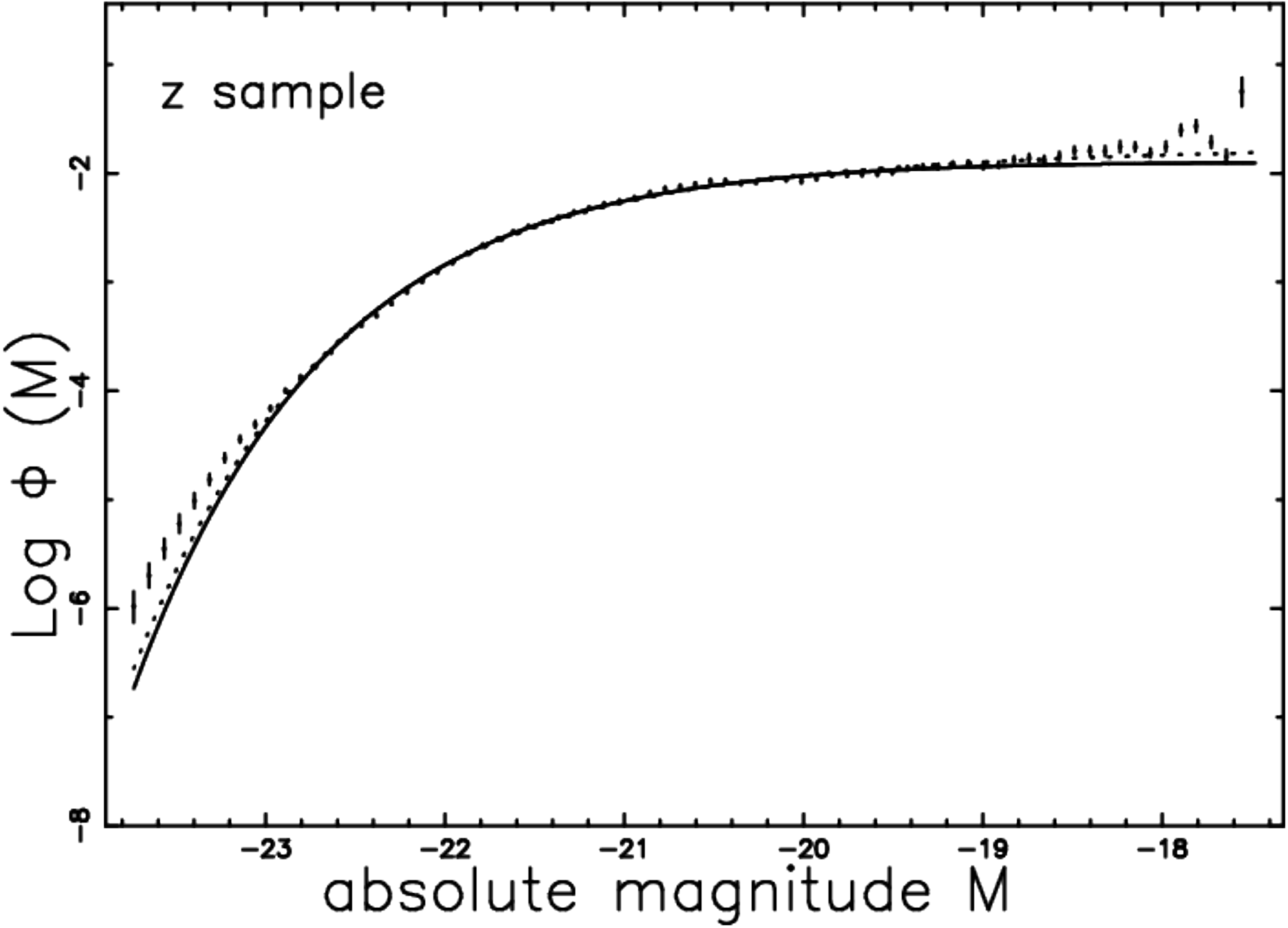}
\caption  {The luminosity function data of
SDSS($z^*$) are represented with error bars.
The continuous line fit represents our truncated gamma  LF
(\ref{lfgtmagni}) with
parameters $M_l$=-23.73, $M_u$=-17.48, $M^*$=-21.1, $\Psi^*=0.04\, Mpc^{-3}$ and $c=0.02$.
The dotted
line represents the Schechter LF
with  parameters $\Phi^*=0.013 \, Mpc^{-3}$ and
$\alpha=-1.07$.
 }
 \label{due_z}
 \end{figure}

\begin{figure}
\begin{center}
\includegraphics[width=10cm]{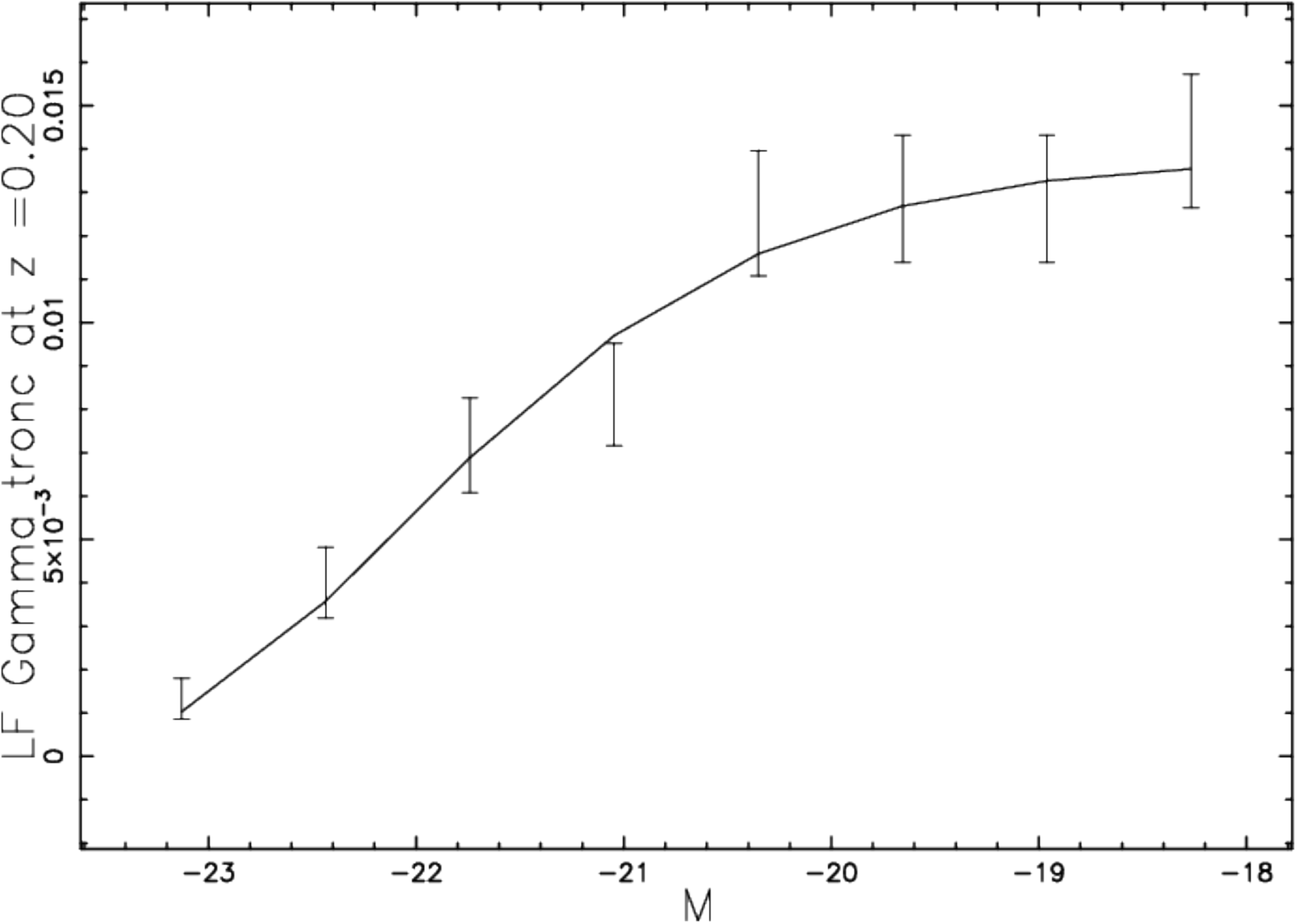}
\end {center}
\caption{
The luminosity function data of zCOSMOS
are  represented with error bars.
The continuous line fit represents our gamma truncated LF
(\ref{lfgtmagni}), the chosen redshift is
$z=0.2$ and $\Delta z$=0.05.
The parameters independent of the redshift
are given in Table \ref{tablezcosmosind}
and the upper magnitude-z relationship is
given in Table \ref{tablezcosmosdep}.
}
          \label{evolution1}%
    \end{figure}

\begin{figure}
\begin{center}
\includegraphics[width=10cm]{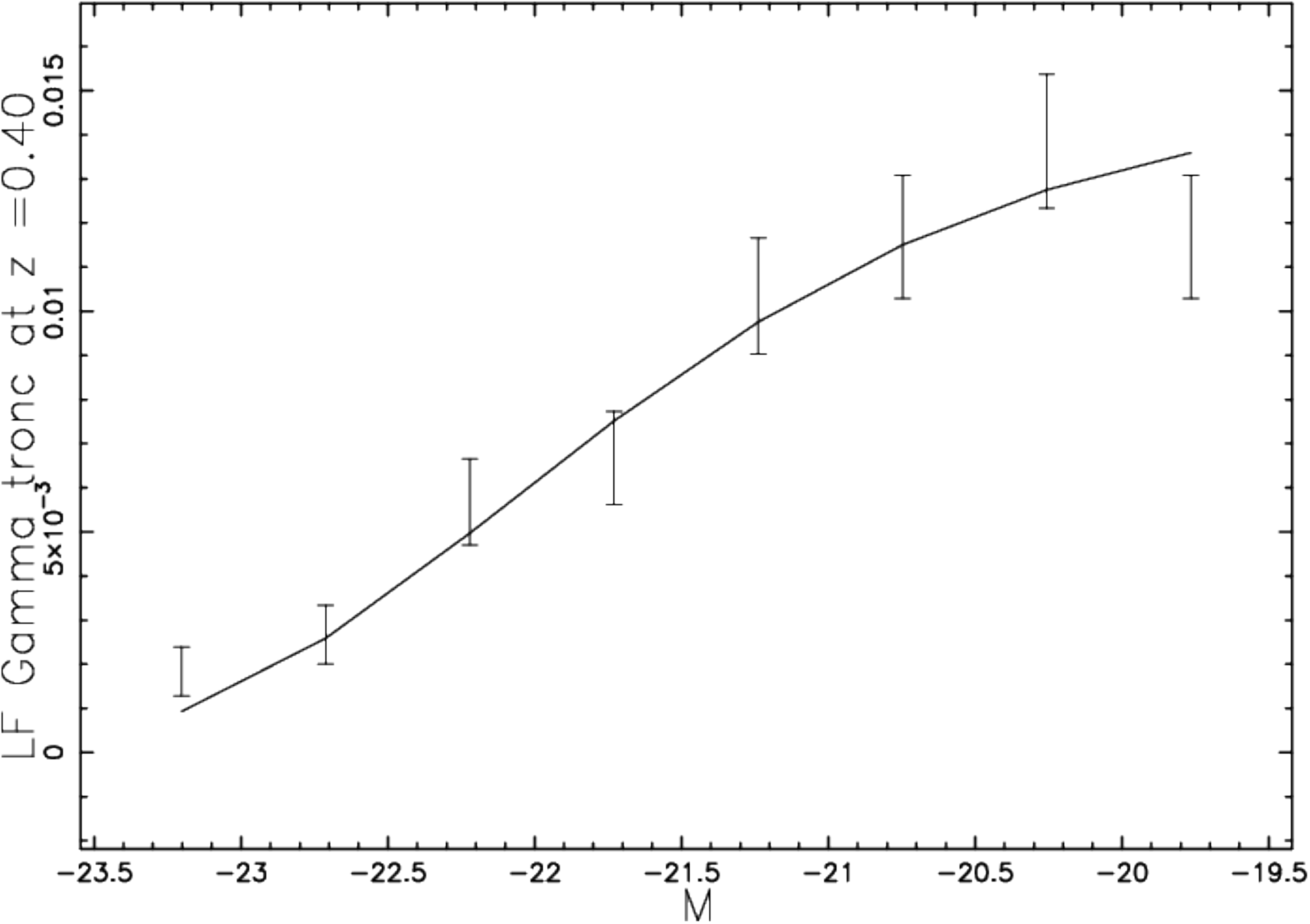}
\end {center}
\caption{
The luminosity function data of zCOSMOS
are  represented with error bars.
The continuous line fit represents our gamma truncated LF
(\ref{lfgtmagni}), the chosen redshift is
$z=0.4$ and $\Delta z$=0.05.
Parameters in Tables \ref{tablezcosmosind}
and \ref{tablezcosmosdep}.
}
          \label{evolution2}%
    \end{figure}

\begin{figure}
\begin{center}
\includegraphics[width=10cm]{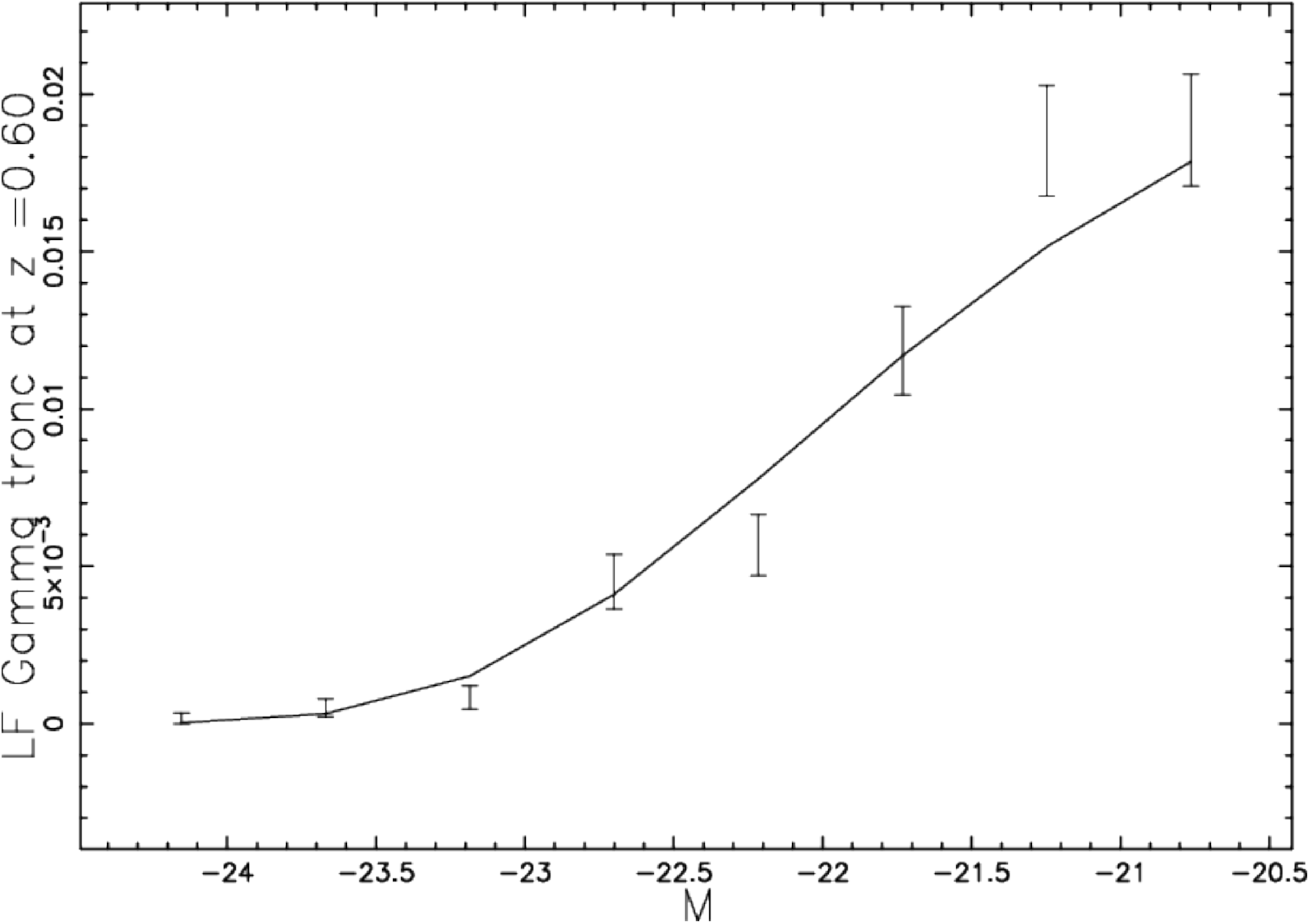}
\end {center}
\caption{
The luminosity function data of
zCOSMOS
are  represented with error bars.
The continuous line fit represents our gamma truncated LF
(\ref{lfgtmagni}), the chosen redshift is
$z=0.6$ and $\Delta z$=0.05.
Parameters in Tables \ref{tablezcosmosind}
and \ref{tablezcosmosdep}.
}
          \label{evolution3}%
    \end{figure}

We assume that the luminosity $L$ takes
values  in the interval
$[L_l, L_u ]$ where the indices $l$ and $u$ mean
lower and upper;
the truncated gamma   LF  is
\begin {equation}
f(L;\Psi^*,L^*,c,L_l,L_u) = \Psi^*\,
k\;\left( {\frac {L}{{\it L^*}}} \right) ^{c-1}{{\rm e}^{-{\frac {L}{{
\it L^*}}}}}
\label{gammatruncated}
\end {equation}
where
$\Psi^*$ is the total number of galaxies per unit Mpc$^3$,
and the constant $k$ is
\begin{equation}
k =
\frac{c}
{
{\it L^*}\, \left(  \left( {\frac {L_{{u}}}{{\it L^*}}} \right) ^{
c}{{\rm e}^{-{\frac {L_{{u}}}{{\it L^*}}}}}-\Gamma  \left( 1+c,{
\frac {L_{{u}}}{{\it L^*}}} \right) +\Gamma  \left( 1+c,{\frac {L_{{
l}}}{{\it L^*}}} \right) - \left( {\frac {L_{{l}}}{{\it L^*}}}
 \right) ^{c}{{\rm e}^{-{\frac {L_{{l}}}{{\it L^*}}}}} \right)
}
\label{constant}
\end {equation}
where
\begin{equation}
\mathop{\Gamma\/}\nolimits\!\left(a,z\right)=\int_{z}^{\infty}t^{{a-1}}e^{{-t}%
}dt
\end{equation}
is the  upper incomplete gamma function,
see
\cite{Abramowitz1965,NIST2010}.
Its expected value   is
\begin{equation}
E(\Psi^*,L^*,c,L_l,L_u)
=\Psi^*
\frac
{
-c \left( \Gamma  \left( 1+c,{\frac {L_{{u}}}{{\it L^*}}} \right) -
\Gamma  \left( 1+c,{\frac {L_{{l}}}{{\it L^*}}} \right)  \right) {
\it L^*}
}
{
\left( {\frac {L_{{u}}}{{\it L^*}}} \right) ^{c}{{\rm e}^{-{\frac {
L_{{u}}}{{\it L^*}}}}}-\Gamma  \left( 1+c,{\frac {L_{{u}}}{{\it
L^*}}} \right) +\Gamma  \left( 1+c,{\frac {L_{{l}}}{{\it L^*}}}
 \right) - \left( {\frac {L_{{l}}}{{\it L^*}}} \right) ^{c}{{\rm e}^
{-{\frac {L_{{l}}}{{\it L^*}}}}}
}
\quad .
\label{meangammatruncated}
\end{equation}
More details on the truncated gamma PDF can be found
in \cite{Zaninetti2013e,Okasha2014}.
The four luminosities
$L,L_l,L^*$ and $L_u$
are  connected with  the
absolute magnitude $M$,
$M_l$, $M_u$ and $M^*$
through the following relationship
\begin{equation}
\frac {L}{L_{\sun}} =
10^{0.4(M_{\sun} - M)}
\, ,
\frac {L_l}{L_{\sun}} =
10^{0.4(M_{\sun} - M_u)}
\,
, \frac {L^*}{L_{\sun}} =
10^{0.4(M_{\sun} - M^*)}
\,
, \frac {L_u}{L_{\sun}} =
10^{0.4(M_{\sun} - M_l)}
\label{magnitudes}
\end{equation}
where the indices $u$ and $l$ are inverted in
the transformation
from luminosity to absolute magnitude
and $M_{\sun}$ is the absolute magnitude
of the sun in the considered band.
The  gamma truncated LF  in magnitude is
\begin{equation}
\label{lfgtmagni}
\Psi (M) dM
=
\frac
{
 0.4\,c \left( {10}^{ 0.4\,{\it M^*}- 0.4\,M} \right) ^{c}{{\rm e}^
{-{10}^{ 0.4\,{\it M^*}- 0.4\,M}}}{\it \Psi^*}\, \left( \ln
 \left( 2 \right) +\ln  \left( 5 \right)  \right)
}
{
D
}
\end{equation}
where
\begin{eqnarray}
D =
{{\rm e}^{-{10}^{- 0.4\,M_{{l}}+ 0.4\,{\it M^*}}}} \left( {10}^{-
 0.4\,M_{{l}}+ 0.4\,{\it M^*}} \right) ^{c}-{{\rm e}^{-{10}^{ 0.4\,
{\it M^*}- 0.4\,M_{{u}}}}} \left( {10}^{ 0.4\,{\it M^*}- 0.4\,M_
{{u}}} \right) ^{c}
\nonumber  \\
-\Gamma  \left( 1+c,{10}^{- 0.4\,M_{{l}}+ 0.4\,{
\it M^*}} \right) +\Gamma  \left( 1+c,{10}^{ 0.4\,{\it M^*}- 0.4
\,M_{{u}}} \right)
\end{eqnarray}
A first  test on the  reliability  of the truncated gamma
LF   was performed on the data
of the  Sloan Digital Sky Survey (SDSS),
see  \cite{Blanton_2003},
in the  band
$z^*$.
The  number of variables  can be reduced to two
once $M_u$ and $M_l$ are identified with
the maximum and  minimum   absolute magnitude of  the
considered sample.
The LFs considered here are displayed in
Figure \ref{due_z}.
%citiamofigura_due_z
% beginning figure due_z
% end figure due_z
A {\it second} test is represented by the behavior
of the LF at high $z$.
We expect a progressive decrease
of
the low luminosity galaxies (high magnitude)
when $z$ is increasing.
A formula which models the previous statement can be obtained
by Eq.~(\ref{absmagz}),
which models the absolute magnitude, $M$,
as a function of the redshift, inserting as the apparent magnitude,
$m$, the
limiting magnitude of the considered catalog.
We now outline how to build an observed LF for a galaxy in
a consistent way; the selected catalog is
zCOSMOS,
which is made up of 9697 galaxies up to $z=$4, see \cite{Lilly2009}.
The observed LF for zCOSMOS can be built by employing the
following algorithm.
\begin{enumerate}
\item The minimax approximation for the luminosity distance
      in the case of the JLA compilation parameters, see Eq.~(\ref{dlzminimaxjla}) allows
      fixing the distance, in the following $r$,
      once $z$ is given.
\item A value for the redshift is fixed, $z$,  as well as the
      thickness of the layer, $\Delta z$.
\item All the galaxies comprised between $z$ and $\Delta z$
are selected.
\item The absolute magnitude is  computed  from Eq.~(\ref{absmagz}).
\item The distribution in magnitude is organized in
  frequencies versus absolute magnitude.
\item The frequencies are divided by the volume,
    which is $V=\Omega  \pi r^2 \Delta r$,
    where $r$ is the considered radius, $\Delta r$ is the thickness
    of the radius, and $\Omega$ is the solid angle of ZCOSMOS.
\item The error in the observed LF is obtained  as
      the square root of the frequencies divided by the
      volume.
\end{enumerate}
Figures  \ref{evolution1},
%citiamofigura_evolution1
\ref{evolution2},
%citiamofigura_evolution2
and
\ref{evolution3}
%citiamofigura_evolution3
present  the LF of zCOOSMOS as well
as the fit with the truncated   beta LF
at $z=0.2$, $z=0.4$, and $z=0.6$, respectively.
%begin figure evolution1
%end  figure evolution1

%begin figure evolution2
%end  figure evolution2

%begin figure evolution3
%end  figure evolution3

\begin{table}
\caption
{
Parameters  of the gamma truncated LF independent of $z$ when
$c=0.01$.
}
 \label{tablezcosmosind}
 \[
 \begin{array}{ccc}
 \hline
 \hline
 \noalign{\smallskip}
M_l  & M^* &  c \\
 \noalign{\smallskip}
 \hline
 -23.47 & -22.7  & 0.01\\
 \hline
 \hline
 \end{array}
 \]
\end{table}

\begin{table}
\caption
{
Upper magnitude, $M_u$\ (mag),  and normalization, $\Psi^*$\ Mpc$^{-3}$,
dependence  on $z$ when $c=0.01$.
}
 \label{tablezcosmosdep}
 \[
 \begin{array}{ccc}
 \hline
 \hline
 z & \Psi^*    & M_u  \\
  \hline
0.2 &  0.0659 & -16.76  \\
  \hline
0.4 &  0.0459& -18.48  \\
  \hline
0.6 &   0.0479 & -19.55  \\
 \hline
 \hline
 \end{array}
 \]
\end{table}

\section{Different Cosmologies}

\label{seccosmologies}
Here we analyse the distance modulus for SNe  in
other  cosmologies
in the framework general relativity (GR),
expanding flat universe,
special relativity (SR) and Euclidean static universe.

\subsection{Simple GR cosmology}

In the framework of GR
the  received flux, f, is
\begin{equation}
f=\frac{L}{4\,\pi d_L^2}
\quad  ,
\end{equation}
where $d_L$ is the luminosity distance which depends from the
cosmological model adopted, see Eq.~(7.21) in
\cite{Ryden2003} or Eq.~(5.235) in \cite{Lang2013}.

The distance modulus in the simple GR cosmology  is
\begin{equation}
m-M =
43.17-{\frac {1}{\ln  \left( 10 \right) }\ln  \left( {\frac {{\it H_0}
}{70}} \right) }+5\,{\frac {\ln  \left( z \right) }{\ln  \left( 10
 \right) }}+ 1.086\, \left( 1-{\it q_0} \right) z
\quad  ,
\label{modulussimple}
\end{equation}
see  Eq.~(7.52) in  \cite{Ryden2003}.
The number of free parameters in the simple GR cosmology
is  two: $H_0$ and $q_0$.

\subsection{Flat expanding universe.}

This model is based on the standard definition
of luminosity in the flat expanding universe.
The luminosity distance, $r_{L}^{\prime}$, is
\begin{equation}
r_L^{\prime} =\frac{c}{H_0} z
\quad ,
\end{equation}
and the distance modulus
is
\begin{equation}
m-M = = -5 \log_{10}  +5 \log_{10}  r_L^{\prime} +2.5 \log(1+z)
\quad  ,
\label{modulusheymann}
\end{equation}
see formulae (13) and (14) in \cite{Heymann2013}.
The number of free parameters in the flat expanding model.
is one: $H_0$.

\subsection{Einstein-De Sitter universe in SR}

In the Einstein--De Sitter model, which is developed in
SR,
the luminosity distance, after \cite{Einstein1932,Krisciunas1993},
is
\begin{equation}
d_L =2\,{\frac {c \left( 1+z-\sqrt {z+1} \right) }{H_{{0}}}}
\quad ,
\end{equation}
and the distance modulus for the Einstein-De Sitter model is
\begin{equation}
m-M =25+5\,{\frac {1}{\ln  \left( 10 \right) }\ln  \left( 2\,{\frac {c
 \left( 1+z-\sqrt {z+1} \right) }{H_{{0}}}} \right) }
\quad .
\label{moduluseds}
\end{equation}
The number of free parameters in the Einstein-De Sitter model
is one: $H_0$.

\subsection{Milne universe in SR}

In the Milne model, which is developed in the framework of SR,
the luminosity distance,
after \cite{Milne1933,2005Chodorowski,Adamek2014}, is
\begin{equation}
d_L ={\frac {c \left( z+\frac{1}{2}\,{z}^{2} \right) }{H_{{0}}}}
\quad ,
\end{equation}
and the distance modulus for the Milne model is
\begin{equation}
m-M =
25+5\,{\frac {1}{\ln  \left( 10 \right) }\ln  \left( {\frac {c \left(
z+\frac{1}{2}\,{z}^{2} \right) }{H_{{0}}}} \right) }
\quad .
\label{modulusmilne}
\end{equation}
The number of free parameters in the Milne model
is one: $H_0$.

\subsection{Plasma cosmology}

In an Euclidean static framework
among many possible absorption mechanisms we selected
a photo-absorption process
between the photon  and the electron in  the IGM.
This relativistic process produces a nonlinear dependence
between redshift and distance
\begin{equation}
z =   \left( \exp (H_0 \, d ) -1 \right)
\quad ,
\label{ashmorez}
\end{equation}
see Eq.~(4) in \cite{Ashmore2006}.
The previous equation is identical to our Eq.~(\ref{nonlzd}).
The Hubble constant in this first plasma model is
\begin{equation}
H_0=1.2649 \, 10^8 \neave  \h0units
\quad ,
\label{h0photoabsortion}
\end{equation}
where $\neave$ is expressed in cgs units.
A second mechanism is a plasma effect which produces the
following relationship
\begin{equation}
d= \frac{c}{H_0} \ln (1+z)
\quad ,
\label{nonlzdari}
\end{equation}
see Eq. (50) in \cite {Brynjolfsson2004}.
Also this second mechanism produces the same nonlinear
d-z dependence
as our Eq.~(\ref{nonlzd}).
In presence of plasma absorption
the observed flux is
\begin{equation}
f = \frac{{L  \cdot \exp \left( { - b  d  -
H_0 d - 2 H_0 d} \right)}}{{4 \pi d^2 }}{\rm{ ,}}
\quad ,
\end{equation}
where the factor $\exp \left( { - b  d } \right)$
is due to  Galactic and host galactic extinctions,
$-H_0 d$ is reduction to the plasma in the IGM and
$- 2 H_0 d$  is the reduction due to Compton scattering,
see formula before Eq.~(51) in \cite {Brynjolfsson2004}.
The resulting distance modulus in the plasma mechanism
is
\begin{equation}
m-M =
5\,{\frac {\ln  \left( \ln  \left( z+1 \right)  \right) }{\ln  \left(
10 \right) }}+\frac{15}{2}\,{\frac {\ln  \left( z+1 \right) }{\ln  \left( 10
 \right) }}+5\,{\frac {1}{\ln  \left( 10 \right) }\ln  \left( {\frac {
c}{H_{{0}}}} \right) }+25+ 1.086\,b
\quad ,
\label{modulusplasma}
\end{equation}
see Eq.~(7) in \cite{Brynjolfsson2006}.
The number of free parameters in the plasma cosmology
is one: $H_0$ when $b=0$.

\subsection{Modified tired light}

\label{subsecplasma}
In an Euclidean static  framework
the modified tired light (MTL)  has been introduced in
Section 2.2 in \cite{Zaninetti2015a}.
The distance in MTL is
\begin{equation}
d= \frac{c}{H_0} \ln (1+z)
\quad .
\label{nonlzd}
\end{equation}
The  distance modulus in the modified tired light (MTL)
is
\begin{equation}
m-M = \frac{5}{2}\,{\frac {\beta\,\ln  \left( z+1 \right) }{\ln  \left( 10 \right) }
}+5\,{\frac {1}{\ln  \left( 10 \right) }\ln  \left( {\frac {\ln
 \left( z+1 \right) c}{H_{{0}}}} \right) }+25
\quad .
\label{modulustired}
\end{equation}
Here $\beta$ is a parameter comprised  between 1 and 3
which allows to match theory with observations.
The  number of free  parameters in MTL
is two: $H_0$ and $\beta$.

\subsection{Results for different cosmologies}

The statistical parameters  for the different    cosmologies
here analysed
can be found in
Table \ref{chi2valuediff} in the case of the Union 2.1 compilation
and in Table \ref{chi2valuediffjla} for the JLA compilation.
\begin{table}[ht!]
\caption
{
Numerical values of  $\chi^2$, $\chi_{red}^2$, $Q$
and the AIC of the Hubble diagram for the Union 2.1 compilation,
$k$ stands for the number of parameters,
$H_0$ is expressed in $\h0units$.
}
\label{chi2valuediff}
\begin{center}
\resizebox{17cm}{!}
{
\begin{tabular}{|c|c|c|c|c|c|c|c|}
\hline
cosmology & Eq. &  k         &   parameters    & $\chi2$& $\chi_{red}^2$ &
Q  & AIC \\
\hline
simple (GR)    & (\ref{modulussimple}) & 2  &$H_0=73.79\pm 0.024$, $q_0$=-0.1
& 689.34 &  1.194  & 8.6 $\,10^{-4}$ &  793.34\\
\hline
flat~expanding~model & (\ref{modulusheymann}) & 1  &
$H_0=66.84\pm 0.22 $
&653  & 1.12 &  0.017  &  655 \\
\hline
Einstein-De Sitter (SR) & (\ref{moduluseds}) & 1  &
$H_0=63.17\pm 0.2 $
& 1171.39 & 2.02& 2 $\,10^{-42}$  &  1173.39\\
\hline
Milne (SR) & (\ref{modulusmilne}) & 1  &
$H_0=67.53 \pm 0.22 $
& 603.37 & 1.04 &  0.23 &  605.37\\
\hline
plasma (Euclidean)      & (\ref{modulusplasma}) & 1 & $H_0=74.2\pm 0.24 $
& 895.53 & 1.546   & 5.2 $\,10^{-16}$  & 897.5 \\
\hline
MTL (Euclidean) & (\ref{modulustired}) & 2 & $\beta$=2.37, $H_0=69.32 \pm 0.34 $  &
567.96 & 0.982 & 0.609 &571.9 \\
\hline
\end{tabular}
}
\end{center}
\end{table}

\begin{table}[ht!]
\caption
{
Numerical values of  $\chi^2$, $\chi_{red}^2$, $Q$
and the AIC of the Hubble diagram for the JLA compilation,
$k$ stands for the number of parameters,
$H_0$ is expressed in $\h0units$.
}
\label{chi2valuediffjla}
\begin{center}
\resizebox{17cm}{!}
{
\begin{tabular}{|c|c|c|c|c|c|c|c|}
\hline
cosmology & Eq. &  k         &   parameters    & $\chi2$& $\chi_{red}^2$ &
Q  & AIC \\
\hline
simple  (GR)  & (\ref{modulussimple}) & 2  &$H_0=73.79\pm 0.023$, $q_0$=-0.14
& 749.14 &   1.016 & 0.369 &  755.14\\
\hline
flat~expanding~model & (\ref{modulusheymann}) & 1  &
$H_0=66.49 \pm 0.18 $
&717.3  & 0.97 &  0.709  &  719.3 \\
\hline
Einstein-De Sitter (SR) &(\ref{moduluseds}) & 1  &
$H_0=62.57\pm 0.17 $
& 1307.75 & 1.76 & 3.27 $\,10^{-34}$  &  1309.75\\
\hline
Milne (SR) &(\ref{modulusmilne}) & 1  &
$H_0=67.19 \pm 0.18 $
& 656.11 & 0.887 &  0.986 &  658.11\\
\hline
plasma    (Euclidean)   & (\ref{modulusplasma})& 1 & $H_0=74.45\pm 0.2$
& 1017.79 & 1.377   & 3.59 $\,10^{-11}$  & 1019.79 \\
\hline
MTL (Euclidean)& (\ref{modulustired}) & 2 & $\beta$=2.36, $H_0=69.096 \pm 0.32$ &
626.27 & 0.848 & 0.998 &630.27\\
\hline
\end{tabular}
}
\end{center}
\end{table}

\section{Conclusions}

{\bf Pad\'e approximant}

It is generally thought
that in the case of the luminosity distance
the  Pad\'e approximant is more accurate than
the Taylor expansion.
As an example, at $z=1.5$, which is the maximum value of the redshift
here considered, the percentage error of the luminosity distance
is
$\delta=0.036\%$ in the case of the Pad\'e approximation.
In the case of of the Taylor expansion, $\delta=0.036\%$
for the luminosity distance
is reached $z=0.322$ which means a more limited range  of
convergence than for the  Pad\'e approximation.
Once a precise approximation for the luminosity
distance was obtained,
see Eq. (\ref{luminositydistance}),
we derived an approximate expression
for the distance modulus,
see Eq. (\ref{distancemodulusexplicit}),
and the absolute magnitude, see Eq.~(\ref{absmagz}).

{\bf Astrophysical Applications}

The availability of the observed distance modulus
for a great number of SNs of  type Ia
allows deducing $H_0$, $\om$ and $\ola$ for two catalogs,
see Table \ref{chi2value}.
In order to derive the above parameters,
the  Levenberg--Marquardt  method was implemented,
and therefore the first derivative of the
distance modulus, see Eq. (\ref{distancemodulusexplicit}),
with respect to three  parameters is provided.
The value of $H_0$ is a matter of research rather than a well
defined constant.
As an example, a recent evaluation with a sample of Cepheids
gives $H_0=73.8 \h0units$, see \cite{Riess2011}.
Once the above value is considered the  `true' value,
we have found, adopting the  Pad\'e  approximant,
$H_0=69.81 \h0units$, which means  a
percentage error $\delta=5.4\%$, for the Union 2.1  compilation
and
$H_0=69.398 \h0units$, which means  a
percentage error $\delta=5.9\%$, for the JLA  compilation,
see Table \ref{chi2value}.

{\bf Evolutionary effects}

The evolution of the LF for galaxies as function of the redshift
is here modeled by an upper and lower truncated gamma PDF.
This choice allows modeling the lower bound in luminosity
(the higher bound in absolute magnitude) according to the
evolution of the absolute magnitude, see Eq.~(\ref{absmagz}).
According to the LF here considered, see Eq.~(\ref{lfgtmagni}),
the evolution with $z$ of the LF is simply connected
with the evolution of the higher bound in absolute magnitude,
see Figures \ref{evolution1}, \ref{evolution2}
and \ref{evolution3}.
Is not necessary to modify the shape parameters of the LF,
which are $c$
and  $M^*$, but only to calculate the normalization $\Psi^*$
at different values of the redshift.

{\bf Statistical tests for Union 2.1 }

In the case of the Union 2.1  compilation,
the best results for $\chi_{red}^2$  are obtained
by the
$\Lambda$CDM cosmology (GR),
$\chi_{red}^2=0.975$,
against
$\chi_{red}^2=0.982$
of the
MTL cosmology (Euclidean),
but
the situation is inverted when the AIC is considered:
the AIC  is 571.9 for the
MTL cosmology and
568.7 for the
$\Lambda$CDM cosmology (GR),
see  Tables \ref{chi2value}
and \ref{chi2valuediff}.

The simple model (GR),
the Einstein--De Sitter model (SR),
the Milne model (SR)
and
the plasma model (Euclidean)
are rejected because
the reduced  merit function $\chi_{red}^2$
is smaller than one,
see Table \ref{chi2valuediff}.
The best performing one-parameter model is that of
Milne, $\chi_{red}^2=1.04 $,
followed by the
flat expanding model,
 $\chi_{red}^2=1.12 $,
 see Table \ref{chi2valuediff}.

{\bf Statistical tests for JLA }

In the case of the JLA compilation,
the best results for $\chi_{red}^2$  are obtained
by the MTL cosmology (Euclidean),
$\chi_{red}^2=0.848$,
against
$\chi_{red}^2=0.849$
for the
$\Lambda$CDM  cosmology (GR),
see  Tables \ref{chi2value}
and \ref{chi2valuediffjla}.
The simple model (GR),
the Einstein--De Sitter model (SR)
and
the plasma model (Euclidean)
are rejected
because the reduced  merit function $\chi_{red}^2$
is smaller than one,
see Table \ref{chi2valuediffjla}.
In the case of the JLA, the  test on the Milne model
is positive because
$\chi_{red}^2$
is smaller than one.
The best performing one-parameter model is that of
Milne, $\chi_{red}^2=0.887$,
followed by the
flat expanding model,
 $\chi_{red}^2=0.97$,
 see Table \ref{chi2valuediffjla}.

{\bf Different Approachs }

Table \ref{problems} reports six  items 
connected with the use of Pad\'e  approximant
in Cosmology: the letter Y/N indicates
if the item  is treated or not and the columns
identifies  the paper in question, 
LF means luminosity function for
galaxies.
%siamoqui
\begin{table}
 \caption {Arguments treated in  papers on Pad\'e approximants  and here }
 \label{problems}
 \[
 \begin{array}{lcccc}
 \hline
 \hline
 \noalign{\smallskip}
Problem
&  Aviles~2014
&  Wei~ 2014
&  Adachi~2012
&  here      \\
 \noalign{\smallskip}
 \hline
 \noalign{\smallskip}
luminosity~distance       &  Y  & Y  &  Y & Y  \\
distance~modulus          &  Y  & Y  &  Y & Y  \\
empty~beam                &  N  & N  &  Y & N  \\
distance~modulus~minimax  &  N  & N  &  N & Y  \\
poles                     &  N  & N  &  N & Y  \\
LF$=f(z)$                   &  N  & N  &  N & Y  \\
\noalign{\smallskip}
\noalign{\smallskip}
  \hline
 \end{array}
 \]
 \end {table}

\appendix
\setcounter{equation}{0}
\renewcommand{\theequation}{\thesection.\arabic{equation}}

\section{The Pad\'e approximant}
\label{appendixa}
Given a function $f(z)$, the Pad\'e  approximant,
after \cite{Pade1892},
is
\begin{equation}
f(z)=\frac{a_{0}+a_{1}z+\dots+a_{p}z^{p}}{b_{0}+b_{1}%
z+\dots+b_{q}z^{q}}
\quad ,
\end{equation}
where the notation is the same as in \cite{NIST2010}.

The coefficients $a_i$ and $b_i$
are found through Wynn's cross rule,
see \cite{Baker1975,Baker1996}
and our choice is $p=2$ and $q=2$.
The choice of  $p$ and $q$ is a compromise between
precision, high values for  $p$ and $q$, and
the simplicity of the expressions to manage,
low values for  $p$ and $q$;
Appendix \ref{appendixb} gives three different
approximations for the indefinite integral for three different
combinations in  $p$ and $q$.
In the case in which $b_0 \neq 0$  we can divide both numerator
and denominator by $b_0$ reducing by one
the number of parameters, see as an example
\cite{Yamada2014}.

The integrand of Eq.~(\ref{integralez}) is
\begin{equation}
\label{argumentint}
\frac{1}{E(z)} =\frac{1}{ \sqrt{\om\,(1+z)^3+\ok\,(1+z)^2+\ola}}
\quad ,
\end{equation}
and the Pad\'e  approximant  gives
\begin{equation}
\label{argumentintpade}
\frac{1}{E(z)} =
\frac{a_0 + a_1 z + a_2 z^2}{b_0 + b_1 z + b_2 z^2}
\quad ,
\end{equation}
where
\begin{eqnarray}
a_0 =
16\, \bigl( 32\,{{\it \ok}}^{3}{\it \ola}+16\,{{\it \ok}}^{2}{{\it
\ola}}^{2}+160\,{{\it \ok}}^{2}{\it \ola}\,{\it \om}+24\,{{\it \ok}}^{
2}{{\it \om}}^{2}+64\,{\it \ok}\,{{\it \ola}}^{2}{\it \om}
\nonumber \\
+320\,{\it
\ok}\,{\it \ola}\,{{\it \om}}^{2}+40\,{\it \ok}\,{{\it \om}}^{3}+96\,{
{\it \ola}}^{2}{{\it \om}}^{2}+
\nonumber \\
192\,{\it \ola}\,{{\it \om}}^{3}
+15\,{{
\it \om}}^{4}\bigr )  \bigl( {\it \om}+{\it \ok}+{\it \ola} \bigr) ^
{4}
\end{eqnarray}
\begin{eqnarray}
a_1 =
4\, \bigr ( 128\,{{\it \ok}}^{4}{\it \ola}+32\,{{\it \ok}}^{3}{{\it
\ola}}^{2}+704\,{{\it \ok}}^{3}{\it \ola}\,{\it \om}
-16\,{{\it \ok}}^{
2}{{\it \ola}}^{2}{\it \om}
\nonumber \\
+1456\,{{\it \ok}}^{2}{\it \ola}\,{{\it \om
}}^{2}+32\,{{\it \ok}}^{2}{{\it \om}}^{3}-64\,{\it \ok}\,{{\it \ola}}^
{3}{\it \om}-384\,{\it \ok}\,{{\it \ola}}^{2}{{\it \om}}^{2}
\nonumber \\
+1512\,{
\it \ok}\,{\it \ola}\,{{\it \om}}^{3}
+50\,{\it \ok}\,{{\it \om}}^{4}-
192\,{{\it \ola}}^{3}{{\it \om}}^{2}-288\,{{\it \ola}}^{2}{{\it \om}}^
{3}+648\,{\it \ola}\,{{\it \om}}^{4}
\nonumber \\
+15\,{{\it \om}}^{5} \bigr )
 \bigl( {\it \om}+{\it \ok}+{\it \ola} \bigr) ^{3}
\end{eqnarray}
\begin{eqnarray}
a_2 =
- \bigl( 256\,{{\it \ok}}^{4}{\it \ola}\,{\it \om}-64\,{{\it \ok}}^{3}
{{\it \ola}}^{3}+320\,{{\it \ok}}^{3}{{\it \ola}}^{2}{\it \om}+960\,{{
\it \ok}}^{3}{\it \ola}\,{{\it \om}}^{2}
\nonumber \\
-320\,{{\it \ok}}^{2}{{\it
\ola}}^{3}{\it \om}+240\,{{\it \ok}}^{2}{{\it \ola}}^{2}{{\it \om}}^{2
}+1440\,{{\it \ok}}^{2}{\it \ola}\,{{\it \om}}^{3}+16\,{{\it \ok}}^{2}
{{\it \om}}^{4}
\nonumber  \\
-1600\,{\it \ok}\,{{\it \ola}}^{3}{{\it \om}}^{2}-480\,
{\it \ok}\,{{\it \ola}}^{2}{{\it \om}}^{3}+1140\,{\it \ok}\,{\it \ola}
\,{{\it \om}}^{4}+20\,{\it \ok}\,{{\it \om}}^{5}
\nonumber \\
-256\,{{\it \ola}}^{4}
{{\it \om}}^{2}
-1600\,{{\it \ola}}^{3}{{\it \om}}^{3}-240\,{{\it \ola}
}^{2}{{\it \om}}^{4}
+380\,{\it \ola}\,{{\it \om}}^{5}
\nonumber \\
+5\,{{\it \om}}^{
6} \bigr)  \bigl( {\it \om}+{\it \ok}+{\it \ola} \bigr) ^{2}
\end{eqnarray}
\begin{eqnarray}
b_0 = 16\, \bigl( {\it \om}+{\it \ok}+{\it \ola} \bigr) ^{9/2} \bigl( 32\,{
{\it \ok}}^{3}{\it \ola}+16\,{{\it \ok}}^{2}{{\it \ola}}^{2}+160\,{{
\it \ok}}^{2}{\it \ola}\,{\it \om}
\nonumber \\
+24\,{{\it \ok}}^{2}{{\it \om}}^{2}+
64\,{\it \ok}\,{{\it \ola}}^{2}{\it \om}+320\,{\it \ok}\,{\it \ola}\,{
{\it \om}}^{2}+40\,{\it \ok}\,{{\it \om}}^{3}+96\,{{\it \ola}}^{2}{{
\it \om}}^{2}
\nonumber \\
+192\,{\it \ola}\,{{\it \om}}^{3}+15\,{{\it \om}}^{4}
 \bigr)
\end{eqnarray}
\begin{eqnarray}
b_1 =
4\, \bigl( {\it \om}+{\it \ok}+{\it \ola} \bigr) ^{7/2} \bigl( 256\,{
{\it \ok}}^{4}{\it \ola}+96\,{{\it \ok}}^{3}{{\it \ola}}^{2}+1536\,{{
\it \ok}}^{3}{\it \ola}\,{\it \om}
\nonumber  \\
+96\,{{\it \ok}}^{3}{{\it \om}}^{2}+
336\,{{\it \ok}}^{2}{{\it \ola}}^{2}{\it \om}+3696\,{{\it \ok}}^{2}{
\it \ola}\,{{\it \om}}^{2}
\nonumber \\
+336\,{{\it \ok}}^{2}{{\it \om}}^{3}-64\,{
\it \ok}\,{{\it \ola}}^{3}{\it \om}+384\,{\it \ok}\,{{\it \ola}}^{2}{{
\it \om}}^{2}+4200\,{\it \ok}\,{\it \ola}\,{{\it \om}}^{3}+350\,{\it
\ok}\,{{\it \om}}^{4}
\nonumber \\
-192\,{{\it \ola}}^{3}{{\it \om}}^{2}
+288\,{{\it
\ola}}^{2}{{\it \om}}^{3}+1800\,{\it \ola}\,{{\it \om}}^{4}+105\,{{
\it \om}}^{5} \bigr)
\end{eqnarray}
\begin{eqnarray}
b_2 =
 \bigl( {\it \om}+{\it \ok}+{\it \ola} \bigr) ^{5/2} \bigl( 512\,{{
\it \ok}}^{5}{\it \ola}+384\,{{\it \ok}}^{4}{{\it \ola}}^{2}+3584\,{{
\it \ok}}^{4}{\it \ola}\,{\it \om}
\nonumber \\
+192\,{{\it \ok}}^{3}{{\it \ola}}^{3
}+1984\,{{\it \ok}}^{3}{{\it \ola}}^{2}{\it \om}+10752\,{{\it \ok}}^{3
}{\it \ola}\,{{\it \om}}^{2}+320\,{{\it \ok}}^{3}{{\it \om}}^{3}
\nonumber \\
+960\,
{{\it \ok}}^{2}{{\it \ola}}^{3}{\it \om}
+5136\,{{\it \ok}}^{2}{{\it
\ola}}^{2}{{\it \om}}^{2}+17760\,{{\it \ok}}^{2}{\it \ola}\,{{\it \om}
}^{3}+840\,{{\it \ok}}^{2}{{\it \om}}^{4}
\nonumber \\
+2752\,{\it \ok}\,{{\it \ola}
}^{3}{{\it \om}}^{2}
+7392\,{\it \ok}\,{{\it \ola}}^{2}{{\it \om}}^{3}+
15060\,{\it \ok}\,{\it \ola}\,{{\it \om}}^{4}+700\,{\it \ok}\,{{\it
\om}}^{5}
\nonumber \\
+256\,{{\it \ola}}^{4}{{\it \om}}^{2}+2752\,{{\it \ola}}^{3}{
{\it \om}}^{3}+3696\,{{\it \ola}}^{2}{{\it \om}}^{4}+5020\,{\it \ola}
\,{{\it \om}}^{5}+175\,{{\it \om}}^{6} \bigr).
\end{eqnarray}

\section{The integrals as functions of $p$ and $q$}
\setcounter{equation}{0}
\label{appendixb}

We now present the indefinite integral
of  (\ref{integralez})
for different values of $p$ and $q$.

In the case $p=1,q=1$,
\begin{equation}
F_{1,1}(z;a_0,a_1,b_0,b_1) =
{\frac {a_{{1}}z}{b_{{1}}}}+{\frac {\ln  \left( zb_{{1}}+b_{{0}}
 \right) a_{{0}}}{b_{{1}}}}-{\frac {\ln  \left( zb_{{1}}+b_{{0}}
 \right) b_{{0}}a_{{1}}}{{b_{{1}}}^{2}}}
\label{integral11}
\end{equation}

In the case $p=2,q=1$,
\begin{eqnarray}
F_{2,1}(z;a_0,a_1,a_2,b_0,b_1) =
1/2\,{\frac {a_{{2}}{z}^{2}}{b_{{1}}}}+{\frac {a_{{1}}z}{b_{{1}}}}-{
\frac {zb_{{0}}a_{{2}}}{{b_{{1}}}^{2}}}+{\frac {\ln  \left( zb_{{1}}+b
_{{0}} \right) a_{{0}}}{b_{{1}}}}
\nonumber \\
-{\frac {\ln  \left( zb_{{1}}+b_{{0}}
 \right) b_{{0}}a_{{1}}}{{b_{{1}}}^{2}}}+{\frac {\ln  \left( zb_{{1}}+
b_{{0}} \right) a_{{2}}{b_{{0}}}^{2}}{{b_{{1}}}^{3}}}
\label{integral21}
\end{eqnarray}

In the case $p=2,q=2$,
\begin{eqnarray}
F_{2,2}(z;a_0,a_1,a_2,b_0,b_1,b_2) =
{\frac {a_{{2}}z}{b_{{2}}}} \nonumber \\
+\frac{1}{2}\,{\frac {\ln  \left( {z}^{2}b_{{2}}+zb
_{{1}}+b_{{0}} \right) a_{{1}}}{b_{{2}}}}-\frac{1}{2}\,{\frac {\ln  \left( {z}
^{2}b_{{2}}+zb_{{1}}+b_{{0}} \right) a_{{2}}b_{{1}}}{{b_{{2}}}^{2}}}
\nonumber \\
+2
\,{\frac {a_{{0}}}{\sqrt {4\,b_{{0}}b_{{2}}-{b_{{1}}}^{2}}}\arctan
 \left( {\frac {2\,zb_{{2}}+b_{{1}}}{\sqrt {4\,b_{{0}}b_{{2}}-{b_{{1}}
}^{2}}}} \right) }
-2\,{\frac {a_{{2}}b_{{0}}}{b_{{2}}\sqrt {4\,b_{{0}}
b_{{2}}
-{b_{{1}}}^{2}}}\arctan \left( {\frac {2\,zb_{{2}}+b_{{1}}}{
\sqrt {4\,b_{{0}}b_{{2}}-{b_{{1}}}^{2}}}} \right) }
\nonumber \\
-{\frac {b_{{1}}a_{
{1}}}{b_{{2}}\sqrt {4\,b_{{0}}b_{{2}}-{b_{{1}}}^{2}}}\arctan \left( {
\frac {2\,zb_{{2}}+b_{{1}}}{\sqrt {4\,b_{{0}}b_{{2}}-{b_{{1}}}^{2}}}}
 \right) }+{\frac {{b_{{1}}}^{2}a_{{2}}}{{b_{{2}}}^{2}\sqrt {4\,b_{{0}
}b_{{2}}-{b_{{1}}}^{2}}}\arctan \left( {\frac {2\,zb_{{2}}+b_{{1}}}{
\sqrt {4\,b_{{0}}b_{{2}}-{b_{{1}}}^{2}}}} \right) }
\label{integral22}
\end{eqnarray}

\section{Minimax  approximation}
\label{appendixc}
\setcounter{equation}{0}

Let $f(x)$ be a real function defined
in the interval
$[a, b]$.  The best  rational approximation of degree $(k, l)$
evaluates the coefficients of the ratio of two polynomials
of degree $k$ and $l$, respectively, which minimizes
the maximum difference of
\begin{equation}
max \bigl |f(x) -\frac{p_{0}+p_{1}x+\dots+p_{k}x^{k}}
{q_0+q_{1}x+\dots+q_{\ell}x^{%
\ell}}\bigr |
\end{equation}
on the interval $[a, b]$.
The quality of the fit is  given by the maximum error
over  the considered range.
The coefficients are evaluated through the Remez algorithm,
see \cite{Remez1934,Remez1957}.
As an example, the minimax of degree (2,2) of
\begin{equation}
f(x) = \frac{\log (1+x)}{x}
\quad ,
\end{equation}
is
\begin{equation}
f(x) = \frac
{
0.206888+ 0.093657\,x+ 0.001573\,{x}^{2}
}
{
0.206895+ 0.196889\,x+ 0.0320939\,{x}^{2}
}
\quad ,
\end{equation}
and the maximum error is
$3.345\,10^{-5}$.
As an example, the minimax rational function approximation
is applied
to the evaluation of the complete elliptic integral of the first
and second kind, see \cite{Fukushima2011}.

%\bibliography{biblio}

\end{document}